\documentclass[preprint,12pt,authoryear]{elsarticle}
\usepackage{graphicx}
\usepackage{epstopdf, epsfig}
\usepackage{subcaption}
\usepackage[section]{placeins}
\usepackage[export]{adjustbox}
\usepackage{gensymb}
\usepackage{amsmath}
\usepackage{dirtytalk}
\usepackage{array,multirow,graphicx}

\usepackage[section]{placeins}
\usepackage{needspace}
\usepackage{color}
\usepackage[utf8]{inputenc}
\usepackage[T1]{fontenc}

\usepackage{siunitx}

\usepackage{amssymb}
\journal{International Journal of Multiphase Flow}

\begin{document}
\emergencystretch 3em
\begin{frontmatter}



\title{Scalar transport and nucleation in quasi-two-dimensional starting jets and puffs}

\affiliation[inst1]{organization={Physics of Fluids Group and Max-Planck, Center for Complex Fluid Dynamics, Faculty of Science and Technology, J.M. Burgers Center for Fluid Dynamics, University of Twente},%Department and Organization
            addressline={}, 
            city={Enschede},
            postcode={7500 AE}, 
            state={},
            country={Netherlands}}

\affiliation[inst2]{organization={Max Planck Institute for Dynamics and Self-Organization},%Department and Organization
            addressline={Am Fa\ss berg 17}, 
            city={G\"{o}ttingen},
            postcode={37077}, 
            state={},
            country={Germany}}
            
\author[inst1]{You-An Lee}
\author[inst1,inst2]{Detlef Lohse}
\author[inst1]{Sander G. Huisman}

\begin{abstract}
We experimentally investigate the early-stage scalar mixing and transport with solvent exchange in a quasi-two-dimensional (quasi-2D) jet. We inject an ethanol/oil mixture upward into quiescent water, forming quasi-2D turbulent buoyant jets and triggering the ouzo effect with initial Reynolds numbers, $Re_0=420,$ $840,$ and $1680$. We study two different modes of fluid supply: continuous injection to study a starting jet and finite volume injection to study a puff. While both modes start with the jet stage, the puff exhibits different characteristics in transport, entrainment, mixing, and nucleation, due to the lack of continuous fluid supply. We also inject a dyed ethanol solution as a passive scalar reference case, such that the effect of nucleation for the ethanol/oil mixture can be disentangled.

For the starting jets, the total nucleated mass from the ouzo mixture seems very similar to that of the passive scalar total mass, indicating a primary nucleation site slightly above the virtual origin above the injection needle, supplying the mass flux like the passive scalar injection. With continuous mixing above the primary nucleation site, the mildly increasing nucleation rate suggests the occurrence of secondary nucleation throughout the entire ouzo jet.  

For the puffs, we show that the puff with the smallest $Re_0$ propagates the fastest and its entrainment lasts the longest. We attribute the superior performance to the buoyancy effect, which transforms a turbulent puff into a turbulent thermal, and has been proven to have stronger entrainment. Although the entrainment and nucleation reduce drastically when the injection stops, the mild mixing still leads to non-zero nucleation rates and the reduced decay of the mean puff concentrations for the ouzo mixture.

Adapting the theoretical framework established in \citet{Landel2012b} for quasi-2D turbulent jets and puffs, we successfully model the transport of the horizontally-integrated concentrations for the passive scalar. The fitted advection and dispersion coefficients are then used to model the transport of the ouzo mixture, from which the spatial-temporal evolution of the nucleation rate can be extracted. The spatial distribution of the nucleation rate sheds new light on the solvent exchange process in transient turbulent jet flows.   
\end{abstract}


\end{frontmatter}



%Chapter 1--------------------------------
%1
\section{Introduction}
The studies of turbulent planar jets \citep{Gutmark1976,Jirka2001} and wakes \citep{Chen1995,Balachandar1999} are highly relevant in geophysics and hydraulic engineering. Typical examples are the turbulent discharge of water streams with sediments and pollutants into quiescent water bodies near estuaries \citep{Fischer1976} and rivers \citep{Uijttewaal2014}. The flows are generally referred to as shallow flows, bounded in one direction by geographical constraints, and thus they can be considered as planar or two-dimensional. \citet{Giger1991} and \citet{Dracos1992} identified that a plane jet starts meandering at a streamwise distance from the origin of ten times the confined water depth, which is defined as the far field. They associated the meandering behavior to the formation of the large-scale coherent structures \citep{Giger1991,Dracos1992}, which have a great influence on the transport and mixing of the flow. 

\citet{Landel2012a} referred to the turbulent jets within the shallow water layers as quasi-2D jets. They \citep{Landel2012a} showed that the quasi-2D jet consists of a fast sinuous core and slow quasi-2D eddies surrounding the core. They observed self-similar Gaussian profiles in both the core and the eddies. The linear growth of the eddies results from the entrainment of ambient fluid. Landel \emph{et al.} extended their efforts from the dynamics of a quasi-2D jet \citep{Landel2012a} to its dispersion and mixing \citep{Landel2012b}. They formulated \citep{Landel2012b} a theoretical framework for the streamwise scalar transport of the horizontally-integrated concentrations, which matched well with their experimental results. The term horizontal refers to the transverse direction. Based on the dominant role of the streamwise transport, the horizontal integration enabled \citet{Landel2012b} to model the scalar transport with a one-dimensional advection-diffusion equation and a mixing-length hypothesis. Their solution is governed by two parameters, namely an advection coefficient $K_a$ and a dispersion coefficient $K_d$, which can be determined by experiments. \citet{Landel2012b} also emphasized that dispersion is responsible for a non-negligible fraction of scalar transport in a confined geometry, which leads to 11$\%$ of the tracers traveling ahead of the advective front. \citet{Rocco2015} applied an analytical framework similar to that in \citet{Landel2012b}, focusing on the buoyancy-dominated plume instead. \citet{Rocco2015} revealed a constant propagation speed $w_e$ scaling with the buoyancy flux $f$ as $w_e \approx 1.3f^{1/3}$. They also calculated the advection and dispersion coefficients, which are both smaller than the corresponding values for a quasi-2D jet obtained in \citet{Landel2012b}. 

While the aforementioned studies addressed the flow with continuous volume influx in the steady state, the scenarios with unsteady driving, namely a finite volume impulse or a starting jet and plume, are more complex. These unsteady scenarios have also attracted much scientific interest. \citet{Diez2003} listed the scaling laws for the bulk properties for all the scenarios, including puffs, thermals, starting jets, and starting plumes. Table \ref{tbl:type1} lists the differences of these scenarios. 

\begin{table}
\begin{center}
\def\arraystretch{2}
\setlength{\tabcolsep}{7pt}
\begin{tabular}{cccc} 
 & &\multicolumn{2}{c}{\textbf{Injection Mode}}  \\
& & \textit{Cont.} & \textit{F.V.}\\
\multirow{2}{*}{\rotatebox[origin=c]{90}{\parbox[c]{1.5cm}{\centering \textbf{Dominant Force}}}}  
& \multirow{1}{*}{\rotatebox[origin=l]{90}{\parbox[c]{0.55cm}{$M$}}} & Jet & Puff\\
&\multirow{1}{*}{\rotatebox[origin=c]{90}{\parbox[c]{0.35cm}{$B$}}} & plume & thermal\\ 
\end{tabular}
\end{center}
\vspace{2mm}
\caption{Types of turbulent jet flows discussed in this chapter, where Cont. is abbreviation for continuous, F.V. for finite volume injection, $B$ for buoyancy, and $M$ for momentum.}
\label{tbl:type1}
\end{table}

With a finite volume impulse, a momentum-dominated puff or a buoyancy-dominated thermal can be generated. Applications of turbulent puffs are in pulsed combustion \citep{Johari1993,Ghaem-Maghami2007}, which were shown to provide faster fuel-air mixing than steady jets. The COVID-19 pandemic \citep{Bourouiba2021} has made studying multiphase turbulent puffs crucial in order to understand the transmission of respiratory diseases \citep{Mazzino2021,Chong2021}. \citet{Mazzino2021} obtained the scaling laws for the bulk properties and the structure functions analytically and numerically. \citet{Chong2021} demonstrated that the mixing of the local relative humidity field determines the lifetime of small respiratory droplets, which agrees with the findings in \citet{Rivas2016, Villermaux2017}. Turbulent thermals have sparked interest in the atmospheric research community to facilitate the understanding of cumulus clouds, such as the effect of humidity on entrainment \citep{Hannah2017,Vybhav2022} and the scaling laws of the bulk properties \citep{Skvortsov2021}. 

A starting jet \citep{Cossali2001,Ai2005} (or plume \citep{Turner1962, Middleton1975}) is in a transient state before reaching a steady jet (or plume) or before the full release of a puff (or thermal). A starting jet (or plume) consists of a head vortex and a trailing jet (or plume), and thus the interaction between these two parts governs the overall dynamics and mixing \citep{Ai2006}.

The discussed research efforts on unsteady turbulent jet flows, namely puffs, thermals, starting jets, and starting plumes, are mostly for 3D flows. These flows in a quasi-2D geometry remain mostly unexplored. One of the few exceptions is the work by \citet{Landel2012b}. Using a rectangular source function, they extended their theoretical framework for a quasi-2D steady jet to a quasi-2D puff. They concluded that there are 50$\%$ of the dye traveling ahead of the advective front.

The ouzo effect, also known as solvent exchange, is a physicochemical hydrodynamical process leading to the nucleation of micro-sized oil droplets \citep{Vitale2003,Lohse2020}. The ouzo effect occurs in a ternary liquid solution, which consists of a good solvent with dissolved solute, and a poor solvent. Introducing the poor solvent into the solution will lower the solubility of the solute, leading to saturation and eventually precipitation of the solute. On one hand, the process has been quantitatively studied in micro-scale and in the laminar regime, from a single droplet system \citep{Tan2019} to microfluidic channels \citep{Zhang2015, Hajian2015, Li2021}. On the other hand, solvent exchange in turbulent flows is a new topic which has received much less attention. Although in \citet{Lee2022} we obtained the time-averaged concentration field of the nucleated oil droplets in a round 3D-jet, the opaque nature of the flow makes it difficult to analyze the temporal fluctuations and the unsteady temporal development for a 3D-jet. The quasi-2D geometry enables us to look into these aspects as the thickness of the light path is uniform across the transverse direction. 

In this paper, we aim to explore the solvent exchange process in unsteady turbulent jet flows, that is, starting jets and puffs. In \S2 we briefly go through the experimental setup and the methods to measure the concentration. To present the results, we start with flow visualization and the qualitative description of the jets and puffs in \S3.1. The front propagation, the mean concentration for the puffs, and the corresponding scaling laws are then discussed in \S3.2. In \S3.3, from the time evolution of the total volume and mass, we show the entrained volume flux and the nucleation rate. Last, we calculate the advection and dispersion coefficients in the models adapted from \citet{Landel2012b} in \S3.4, which leads to the estimation for the nucleated oil concentrations as a function of time, position, and the initial Reynolds number. Conclusions are drawn in \S4.   

%-----------------------------------
%2
\section{Experiment}
\subsection{Set-up}
%2.1
\begin{figure}[h!bt]
\centering
\includegraphics[scale=1]{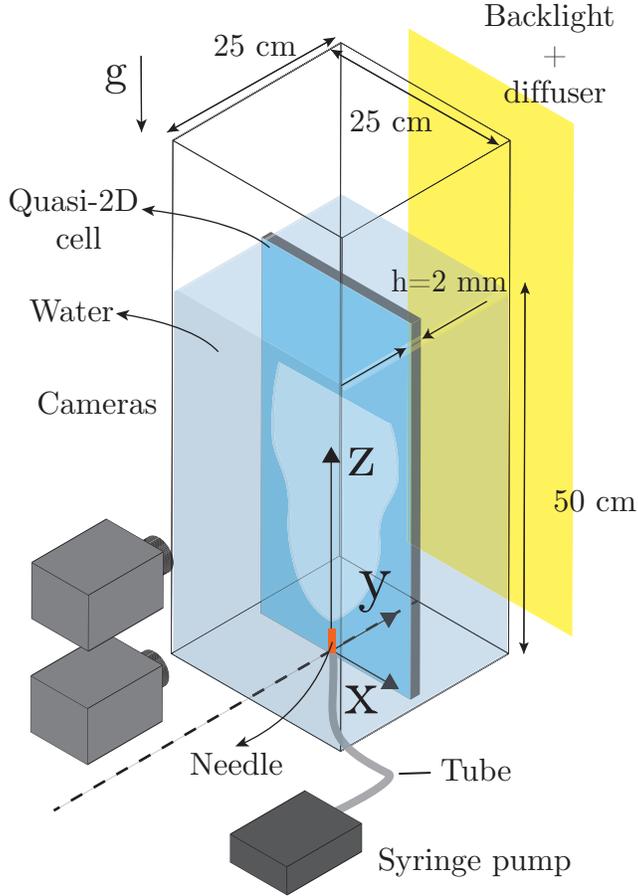}
%\vspace{-5mm}
\caption{Experimental set-up. The syringe pump injects the ethanol/oil mixture upwards to a quasi-2D geometry immersed in a water tank.}
\label{img:setup3}
\end{figure}

We have performed experiments in a water tank filled with decalcified water with dimensions \SI{25}{\cm} $\times$ \SI{25}{\cm} $\times$ \SI{50}{\cm} (W $\times$ L $\times$ H), see Fig.\ \ref{img:setup3}. We injected an ouzo mixture of ethanol and trans-anethole (Sigma Aldrich, $\geq$99\%) at a fixed weight ratio $w_e:w_o=100:1$ upwards into the tank, forming a turbulent buoyant jet. The mixture was injected through a round needle with an inner diameter d = \SI{0.84}{\mm} and a length $\ell$ = \SI{25.4}{\mm}. Two pieces of glass were inserted into the tank to confine the jet to a gap of \SI{2}{\mm} (quasi-2D, thin-cell confinement). The ouzo mixture was injected by a Harvard 2000 syringe pump at three different source flow rates $Q_\text{0}$ to reach three initial Reynolds numbers, $Re_0=u_0d/\nu=420,$ $840$, and $1680$. To produce turbulent puffs, we used the finite volume injection mode of the pump to inject a fixed volume of \SI{1}{\ml} for all three $Re_0$. The water in the tank and the mixture to be injected were kept at $\SI{20}{\celsius}\pm \SI{1}{\kelvin}$ so that the temperature dependence of the solubility is kept to a minimum. We conducted the reference experiments using \SI{7000}ppm dyed ethanol with red food dye (JO-LA) as ethanol is also the dominant component ($>$ 99\%) of the ouzo mixture.

Using known concentrations of ouzo mixture and reference dyed ethanol, we measured the light attenuation, forming a calibration curve by performing an in-situ calibration. The technique requires a backlit optical setting shown in Fig.\ \ref{img:setup3}. We measured the degree of light attenuation using two Photron FASTCAM Mini AX200 high-speed cameras with Zeiss \SI{100}{\mm} focal length objectives. The images were recorded with a 1024 $\times$ 1024 pixels resolution at 50 fps. The cameras were installed at two axial positions to capture the motion and the concentration fields of the jets and puffs up to \SI{25}{\cm} above the injection point. We carefully arranged the two cameras to ensure more than 10$\%$ of the overlap between the two recordings. The experiment for each condition is repeated 10 times for reliable statistical results. In the analysis, we first obtain the desired quantities as a function of time and distance for each experiment, followed by an ensemble average of 10 experiments.  

%-----------------------------------
%2.2
\subsection{Oversaturation}
\begin{figure}[h!bt]
\centering
\includegraphics[scale=1]{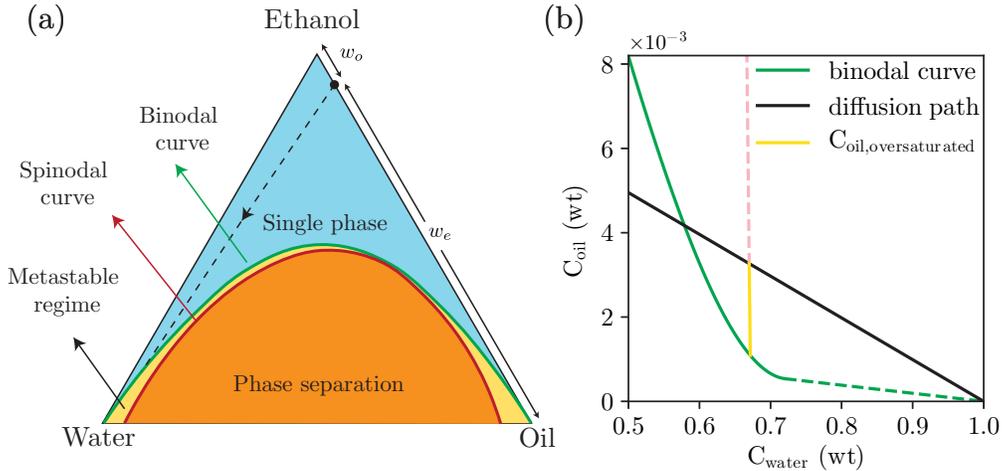}
\caption{Schematic ternary phase diagram and estimation of oversaturation. (a)Ternary phase diagram. The metastable regime in (a) is where the ouzo effect occurs. (b) Binodal curve and the theoretical diffusion path for ouzo mixture ${w_e/w_o = 100}$. The black dashed line in (a) corresponds to the black diffusion path in (b). The binodal curve in (b) is determined partly by titration experiments (green solid line) and partly by direct linear approximation (green dashed line). The length of the yellow line segment in (b) measures the oversaturation. The pink dashed line in (b) extends from the yellow line segment to the pure oil phase, which can be approximated with a vertical line considering the tiny amount of oil.}
\label{img:phasegram}
\end{figure}

To measure the concentration of the nucleated oil, we need to estimate its oversaturation in the ternary liquid solution. Fig.\ \ref{img:phasegram}(a) shows the ternary phase diagram. The metatable regime colored in yellow marks the region where the ouzo effect appears. Following the black dashed line, the initial ethanol/oil mixture approaches the pure water phase as the local fluid parcel mixes with more water. When the local water fraction reaches a certain level, the black dashed line meets the green binodal curve, where the oil becomes saturated and starts to nucleate. 

To simplify the interpretation, the three-component diagram in Fig.\ \ref{img:phasegram}(a) is converted into two components in Fig.\ \ref{img:phasegram}(b). The black dashed line in Fig.\ \ref{img:phasegram}(a) is the so-called diffusion path seen in Fig.\ \ref{img:phasegram}(b), approximating the phase trajectory upon mixing. The difference between the diffusion path and the binodal curve is the estimated oversaturation at a given water fraction, see the yellow line segment in Fig.\ \ref{img:phasegram}(b).   

Knowing the relation between the water fraction and the oversaturation, we can then obtain in-situ calibration curves in Appendix A (Fig.\ \ref{img:calicurve3}) to convert the recorded light intensity fields to the oversaturation fields. For the reference dye case, a different set of calibration curves can convert the light intensity fields to the concentration fields. See Appendix A and Fig.\ \ref{img:calicurve3} for more details. 
%-----------------------------------
%3
\section{Analysis and Results}
%-----------------------------------
%3.1
\subsection{Flow visualization}

\begin{figure}[h!bt]
\centering
\includegraphics[scale=1]{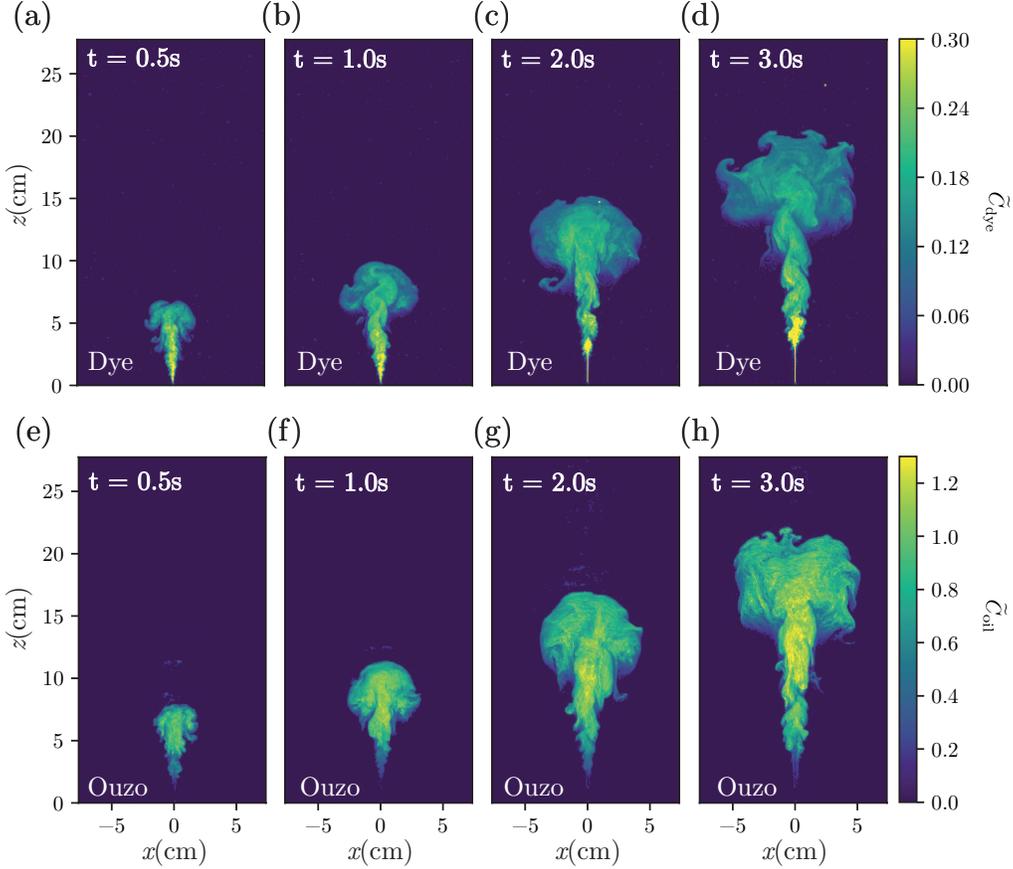}
\caption{The propagation of the starting jet for $Re_0=840$ at four time steps for (a--d) the dye case and (e--h) the ouzo case. The color bars show the corresponding normalized concentrations. Scale bars in (a) and (e) are for (a--d) and (e--h), respectively.}
\label{img:snapjet}
\end{figure}

\begin{figure}[h!bt]
\centering
\includegraphics[scale=1]{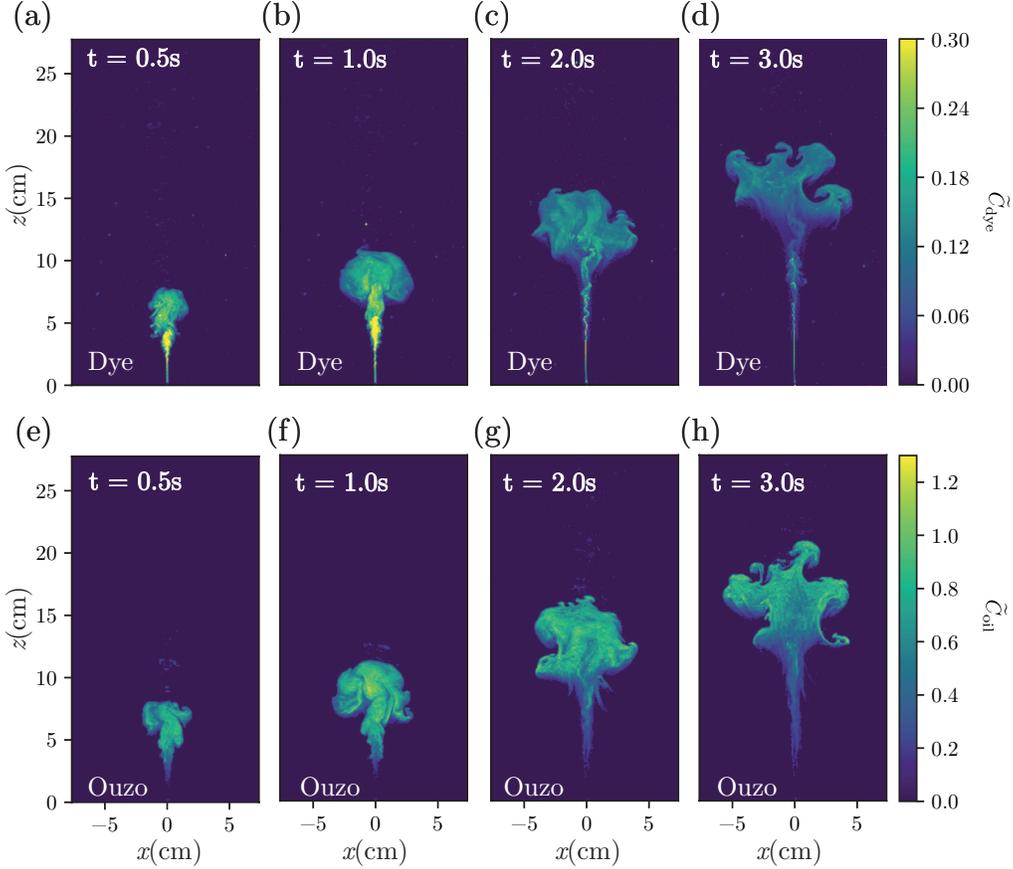}
\caption{The propagation of the puff for $Re_0=840$ at four time steps for (a--d) the dye case and (e--h) the ouzo case. The color bars show the corresponding normalized concentrations.}
\label{img:snappuff}
\end{figure}

We have recorded the starting jets and the puffs for both ouzo mixture and dyed ethanol for each $Re_0$. With three different $Re_0$, we have in total $2\times2\times3=12$ sets of experiments. To provide a qualitative description of the flow before diving into the analysis, we present the concentration fields for $Re_0=840$ at four instances for the starting jets in Fig.\ \ref{img:snapjet}, and for the puffs in Fig.\ \ref{img:snappuff}. 

From Figs.\ \ref{img:snapjet} and \ref{img:snappuff}, we can clearly identify where the flows turn turbulent, which will be referred to as the virtual origin. The distance between the virtual origin and the needle tip is called jet laminar length \citep{Hassanzadeh2021}. This distance and the corresponding time traveling to the virtual origin are subtracted in the following analysis. The normalized concentrations displayed in the color code are defined as:

\begin{align}
& \widetilde{C}_{\text{dye}} = \frac{C_{\text{dye}}}{C_0}, \label{eq:cnormdye}\\
& \widetilde{C}_{\text{oil}} = \frac{C_{\text{oil,sat}}-C_{\text{thres}}}{C_{\text{max}}-C_{\text{thres}}}, \label{eq:cnormoil}
\end{align}

\noindent where $C_0$ is the initial dye concentration \SI{7000}ppm, $C_{\text{max}}$ the theoretical maximum oversaturation discussed in \S2.2, and $C_{\text{thres}}$ is the threshold oversaturation to trigger nucleation, which can be explained by the calibration curve in Fig.\ \ref{img:calicurve3}(b).

The starting jets in Fig.\ \ref{img:snapjet} show a pronounced vortex-ring-like head \citep{Maxworthy1974,Glezer1988,Shariff1992} at the jet front, which was not reported in the previous works on quasi-2D jets or plumes \citep{Landel2012b, Rocco2015}. Also, in Figs.\ \ref {img:snapjet}(d,h), instability patterns are formed at the jet front. We attribute the distinct shape at the front to the large density difference between ethanol and water, triggering a Rayleigh-Taylor instability. Such a clear distinction between the vortex-ring-like head and the trailing jet reminds us of the structure of a starting 3D forced plume \citep{Ai2006}, which becomes the starting point for our modelling attempt in \S3.4.

Comparing the starting ouzo jet in Figs.\ \ref {img:snapjet}(e--h) to the dye jet in Figs.\ \ref {img:snapjet}(a--d), the ouzo jet only turns visible above the virtual origin, where the jet becomes turbulent. Moreover, while the dyed ethanol jet gets diluted as its front propagates, the ouzo jet hardly gets diluted, and its concentration peaks between the head and the trailing jet. 

The puffs in Fig.\ \ref{img:snappuff} are actually starting jets before the injection stops, see Figs.\ \ref {img:snappuff}(a,b,e,f). When the whole \SI{1}{\ml} of the dyed ethanol or the ouzo mixture is fully injected, the puff propagates downstream with a weak tail. The tail is produced by detrainment \citep{McKim2020,Vybhav2022} from the puff and the delayed injection from the pump, which we do not specifically address in this study. Without a continuous supply of the jet fluid from the trailing jets, the puffs exhibit lower concentrations and more pronounced instability patterns, see Figs.\ \ref {img:snappuff}(c,d,g,h).

 %-----------------------------------
%3.2
\subsection{Front and concentration propagation}

\citet{Diez2003} presented the scaling laws for the streamwise penetration as a function of time for 3D puffs, thermals, starting jets, and starting plumes. Utilizing the scaling laws of the velocity evolution for planar jets, plumes, puffs, and thermals, the theoretical scaling exponents for the front penetration can be obtained. Here we look at the experimental front penetration in the quasi-2D cases, aiming to reveal the difference in dynamics between the starting jets and the puffs. 

\begin{figure}[h!bt]
\centering
\includegraphics[scale=1.1]{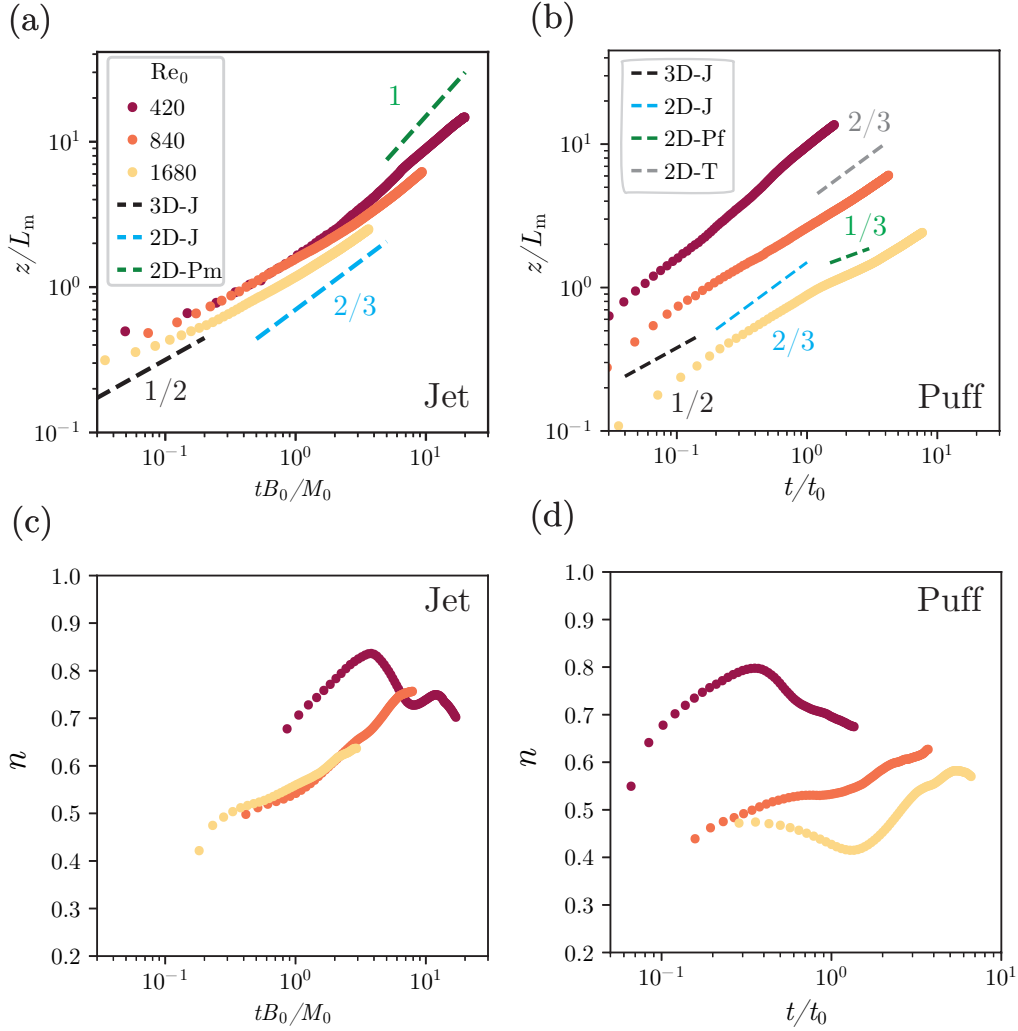}
\caption{Front propagation (a,b) and the corresponding local scaling exponent (c,d) for the starting dye jets (a,c) and the dye puffs (b,d). $L_m = M_0/B_0^{2/3}$ is the momentum length scale for 2D flows, where $M_0$ and $B_0$ are the 2D momentum and buoyancy flux, respectively. $M_0/B_0$ is the corresponding 2D momentum time scale, $t_0$ denotes the duration of injection for the puffs, and $n$ is the local scaling exponent for the front propagation in time, i.e. $n=\frac{\Delta \ln{z/L_m}}{\Delta \ln{tB_0/M_0}}$ for (a,c) and $n=\frac{\Delta \ln{z/L_m}}{\Delta \ln{t_0}}$ for (b,d). The color dashed lines in (a,b) mark the theoretical scaling exponent for the underlying flow regimes, where 3D/2D denotes the three-dimensional and two-dimensional flow, J for jet, Pm for plume, Pf for puff, and T for thermal.}
\label{img:zf}
\end{figure}

As in the developing 3D starting forced plumes detailed in \citet{Ai2006}, our jets and puffs also undergo regime transitions, from jet to plume and from puff to thermal. Upon injection, the jet does not feel the wall yet ($d<h$), so the front propagates as a 3D-jet. As the jet expands, the flow evolves to a quasi-2D jet (or puff). Eventually, when momentum gets depleted, a quasi-2D jet (or puff) transits to a quasi-2D plume (or thermal). Therefore, we expect to see different penetration rates as a function of time for the different cases, which is indeed the case in Figs.\ \ref {img:zf}(a,b). For the jet cases in Figs.\ \ref {img:zf}(a,c), we normalize the distance and the time with the characteristic scales to distinguish the momentum-dominated and the buoyancy-dominated regime. For all three $Re_0$ cases we have, the flows lie in the transitional regime from the jet to the plume, with the $Re_0=420$ jet probing the plume regime deeper, as indicated by the local scaling exponent $n$ in Fig.\ \ref{img:zf}(c), where $z \propto t^n$. 

In Figs.\ \ref {img:zf}(b,d), we normalize the traveling time with the injection duration $t_0$. When the injection stops, we can identify from Fig.\ \ref{img:zf}(d) abrupt transitions in the propagating dynamics. The local scaling exponent $n$ is also smaller than the corresponding jet cases in Fig.\ \ref{img:zf}(c), suggesting a significant effect of the trailing jet on the dynamics.  

We only present the results for the dyed ethanol cases here because we presume that the velocity field of the ouzo jet is very similar to that of the dye case. This is a valid presumption except for the very early stage, see Appendix B and Fig.\ \ref{img:frontcomp} for a detailed comparison.

\begin{figure}[h!bt]
\centering
\includegraphics[scale=1.1]{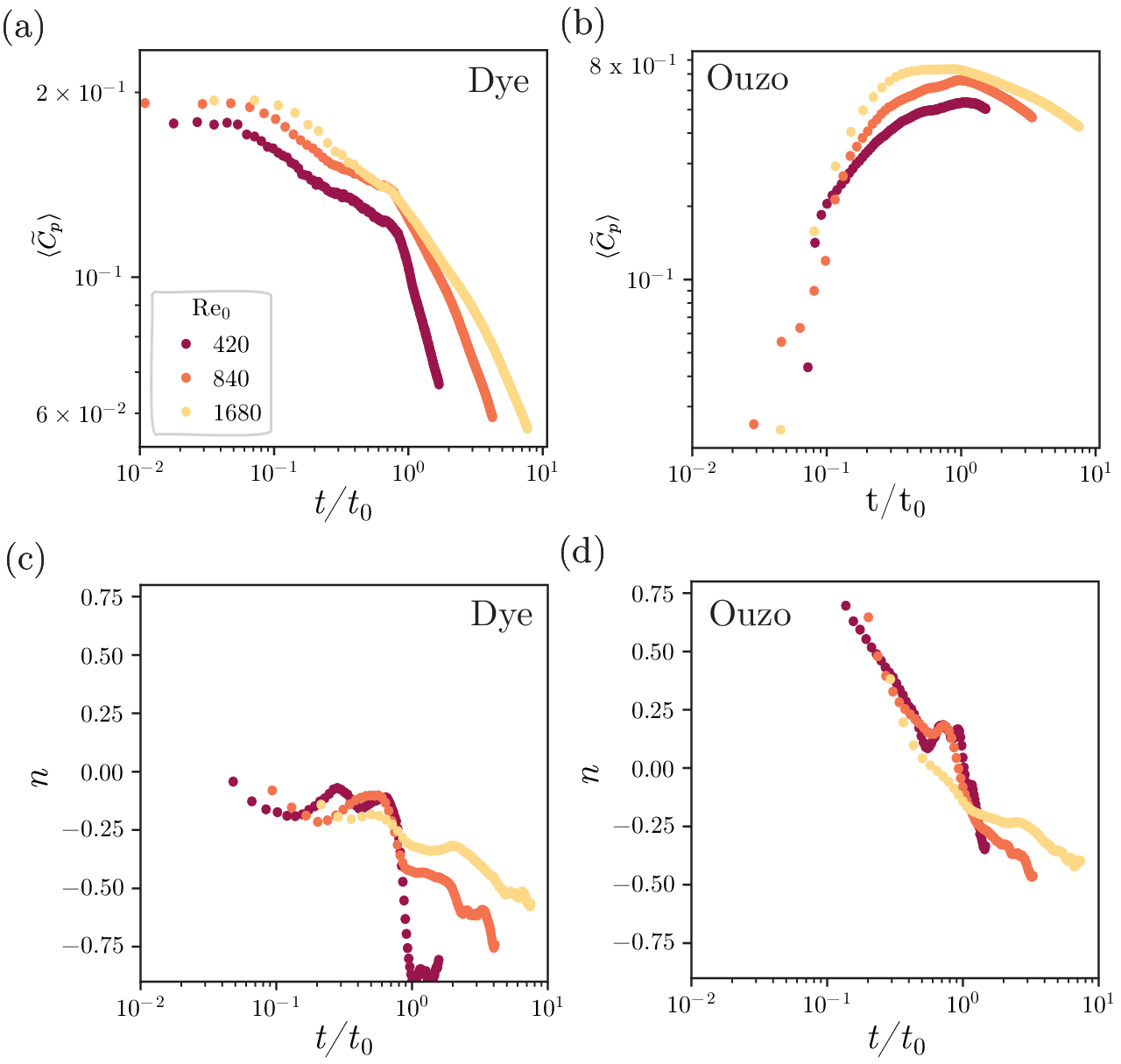}
\caption{Normalized mean concentration evolution for (a) the dye puffs and (b) the ouzo puffs. (c,d) shows the effective scaling exponent (local slope) for (a,b), respectively.}
\label{img:cmeanpuff}
\end{figure}

While we have qualitatively described the concentration evolution of the puffs in \S3.1 and Fig.\ \ref{img:snappuff}, the evolution of the mean concentration can contribute to the quantitative understanding of nucleation in turbulent puffs, see Fig.\ \ref{img:cmeanpuff}. When $t/t_0<1$, a puff behaves as a starting jet. For the dye case, in spite of the continuous supply of the dyed fluid, entrainment and dilution follow the continuous injection, which leads to decreasing concentrations in Fig.\ \ref{img:cmeanpuff}(a). Entering the puff regime with $t/t_0>1$, the concentrations decay faster without a supply of the less diluted fluid. The effective scaling exponent $n$ in Fig.\ \ref{img:cmeanpuff}(c), i.e. the local slope of the double-logarithmic plot, provides clearer information about the change rate of the puff concentration $\langle \widetilde{C}_p \rangle$, which follows an effective scaling law $\langle \widetilde{C}_p \rangle \propto (t/t_0)^n$. While $n$ is almost independent of $Re_0$ in the jet regime, the concentration decays slower with increasing $Re_0$ in the puff regime.   

The oil nucleates in the ouzo jets, which competes with dilution and results in increasing concentrations for $t/t_0<1$ in Fig.\ \ref{img:cmeanpuff}(b). Fig.\ \ref{img:cmeanpuff}(d) shows that $\langle \widetilde{C}_p \rangle$ starts to decrease upon entering the puff regime, $t/t_0=1$ for $Re_0=420,840$, while dilution overcomes nucleation in the jet regime for the highest $Re_0$. From Figs.\ \ref {img:cmeanpuff}(c,d), we see that the ouzo puff decays milder than the corresponding dye puff in the puff regime, which suggests ongoing nucleation even when the injection stops. 

%3.3
\subsection{The entrained volume flux and the nucleation rate}
\begin{figure}[h!bt]
\centering
\includegraphics[scale=1.1]{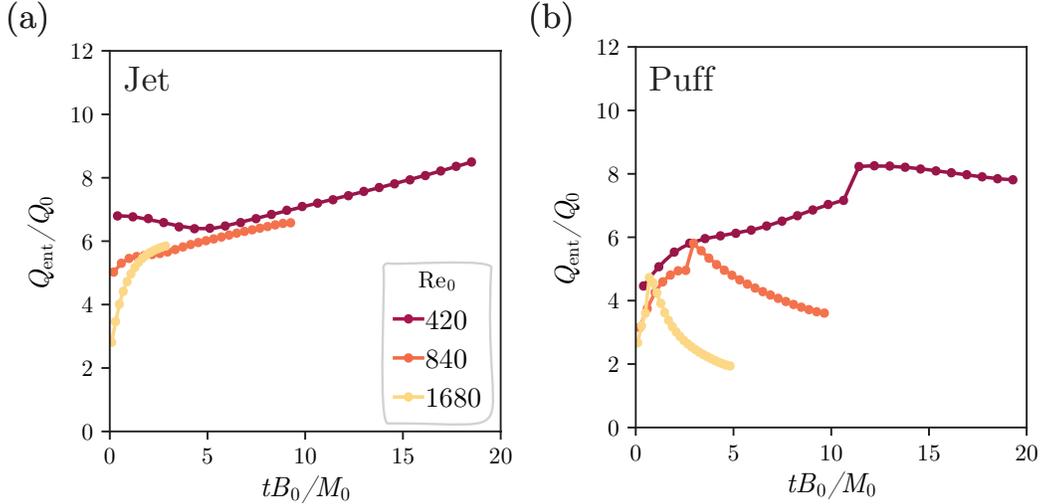}
\caption{The evolution of the normalized entrained volume flux of the dye case for (a) the starting jets, and (b) the puffs.}
\label{img:qent}
\end{figure}

As nucleation of the oil is triggered by entrainment and mixing with ambient water, we focus on the volumetric entrainment rate and the mass evolution in this section. Knowing the total volume of the jet and the puff for each frame, the volume flux $Q$ can be easily obtained by taking the time derivative. Because of the non-monotonic variation of the total volume, we bin the volume evolution using a Gaussian filter. Subtracting the source volume flux $Q_0$ for each $Re_0$ case, the entrained volume flux $Q_{\text{ent}}$ in Fig.\ \ref{img:qent} is calculated. Note that we did not perform the subtraction for the puff cases after the injection stopped. Since the entrainment depends on the momentum and buoyancy regime transition \citep{Fischer1979,Ai2006}, we normalized the traveling time with the momentum length scale discussed in Fig.\ \ref{img:zf}(a). 

\begin{figure}[h!bt]
\centering
\hspace{-5mm}
\includegraphics[scale=1.25]{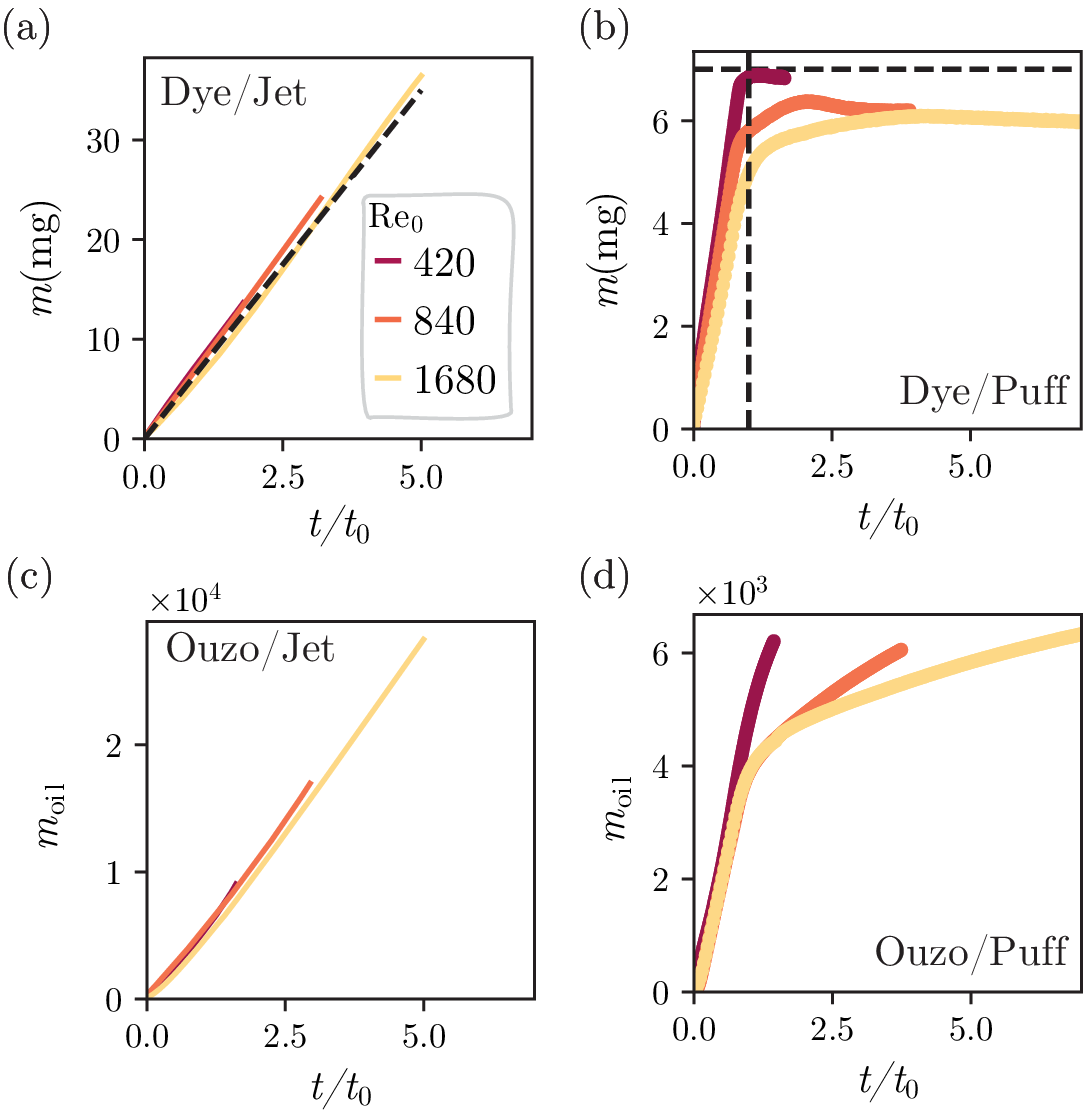}
\caption{The evolution of the total mass for (a,b) the dye case, and (c,d) the ouzo case. The dashed line in (a) marks the theoretical mass evolution of the injected dye, and those in (b) mark the theoretical total mass of the injected dye and the end of injection. Although the starting jets in (a,c) do not have injection duration $t_0$, we use the $t_0$ in the corresponding puffs in (b,d) to normalize the time. Note that the $m_{oil}$ in the ouzo case (c,d) doesn't have a unit because it is not the true mass, but the summation of the normalized concentration in the whole jet, $\iint_A \widetilde{C}_{oil} \,dA$, where A denotes the entire surface area of the jet and $\widetilde{C}_{oil}$ the local normalized oversaturation (oil concentration)}
\label{img:csum}
\end{figure}

Fig.\ \ref{img:qent}(a) shows that the normalized entrained volume flux is almost independent of $Re_0$, which is expected for turbulent jets. The smaller the $Re_0$, the deeper into the plume regime the flow probes, and in turn the higher the normalized entrainment flux. The peaks for all three curves in Fig.\ \ref{img:qent}(b) mark the moment the injection stops, after which the puff phase starts. The normalized entrainment flux for puff decreases with increasing $Re_0$, which is directly related to the moment the injection stops. In other words, the smaller the $Re_0$, the closer the flow approaches the buoyancy-dominated regime, leading to the transition from a plume to a thermal, which entrains more than the jet-to-puff scenario.

While the total volume is dominated by entrainment, the total mass for the dye case accumulates linearly in time with a given mass flux, see Figs.\ \ref {img:csum}(a,b). For the jets, the measured mass evolutions for all three $Re_0$ show excellent agreement with the theoretical value in Fig.\ \ref{img:csum}(a). The puffs, however, do not really reach the intended total mass in Fig.\ \ref{img:csum}(b), especially the two cases with higher $Re_0$. We attribute the deviation to the intense mixing and dilution in the initial period $t/t_0<1$, which might cause a very non-homogeneous distribution of the dye across the gap, compromising the way we estimate the total mass. Also, it is highly likely that there is a small fraction of the fluid that gets too diluted to be detected by the camera, leading to the deviation of the total mass in the final stage.

Looking at the mass evolution for the ouzo jets in Fig.\ \ref{img:csum}(c), the almost-linear curves suggest that most of the oil droplets are generated at a primary nucleation site above the virtual origin, which is essentially like the injection of a normal dyed fluid. The minor increase in the slope of the curves indicates the continuous secondary nucleation along the propagating jets. The ouzo puffs in Fig.\ \ref{img:csum}(d) exhibit pronounced secondary nucleation after stopping the supply from the primary nucleation, coherent with the continuous entrainment reported in Fig.\ \ref{img:qent}(b). Note that the $m_{oil}$ in the ouzo case is not the true mass, but the summation of the normalized concentration in the whole jet, $\iint_A \widetilde{C}_{oil} \,dA$, where A denotes the entire surface area of the jet and $\widetilde{C}_{oil}$ the local normalized oversaturation (oil concentration). Since we can't measure the concentration fields for both nucleated oil and ethanol simultaneously, the lack of local density fields determined by the ethanol/water ratio makes the true oil mass inaccessible.

\begin{figure}[h!bt]
\centering
\includegraphics[scale=1.2]{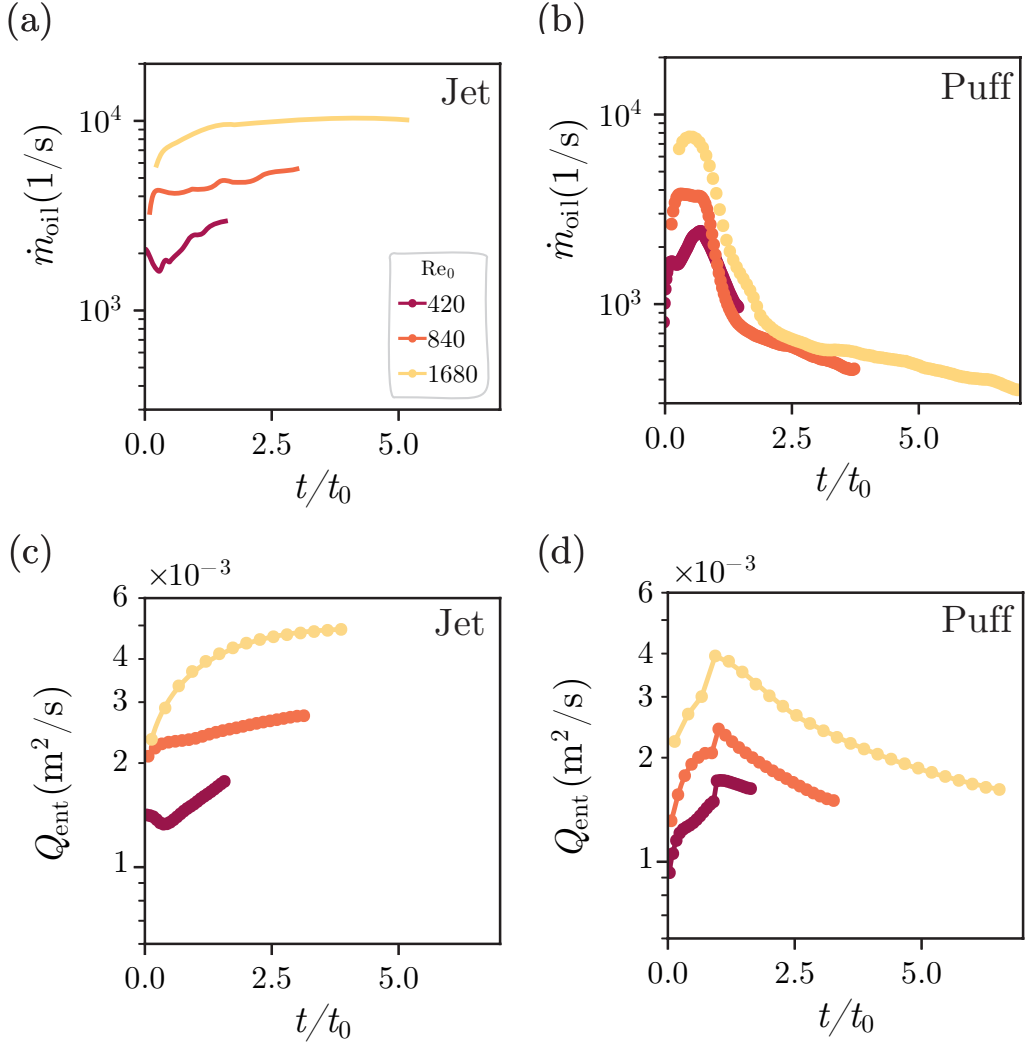}
\caption{The evolution of the mass change rate for (a) the starting ouzo jets, and for (b) the ouzo puffs, and the evolution of the entrainment rate for (c) the starting dye jets (d) for the dye puffs. The mass change rate can be interpreted as the nucleation rate.} 
\label{img:cslope}
\end{figure}

Extracting the slope in Figs.\ \ref {img:csum}(c,d), we estimate the nucleation rate for the ouzo jets and puffs, see Figs.\ \ref {img:cslope}(a,b). The nucleation rate further strengthens our arguments about primary and secondary nucleation. The initial strength of entrainment and mixing, characterized by $Re_0$, dictates the nucleation rate. The plateau for the two cases with higher $Re_0$ in Fig.\ \ref{img:cslope}(a) reveals that most of the oil droplets nucleate at the primary nucleation site, while the $Re_0=420$ case preserves sufficient oil which nucleates downstream with the propagating jet. For the ouzo puffs in Fig.\ \ref{img:cslope}(b), in spite of the drastic decay without the primary nucleation supply for $t/t_0>1$, the nucleation rate for all three cases remains positive. Without the intense mixing and the primary nucleation in the trailing jet, the puff head continuously entrains ambient water, leading to continued nucleation.  

We emphasize that it is entrainment that triggers turbulent mixing and subsequent nucleation of the oil droplets. Therefore, we expect the evolution of the nucleation rate to be more or less affected by that of the entrainment rate. We plot the entrained volume flux in Figs.\ \ref {img:cslope}(c,d), which is the non-normalized version of Fig.\ \ref{img:qent}. Comparing Figs.\ \ref {img:cslope}(a,b) to Figs.\ \ref {img:cslope}(c,d), the evolution of the nucleation rate is strikingly similar to that of the entrained volume flux. The resemblance shows clearly the dominant role of entrainment in triggering nucleation from a global perspective. 

%3.4
\subsection{Transport and nucleation of the horizontally integrated concentrations}

\begin{figure}[h!bt]
\centering
\includegraphics[scale=1.2]{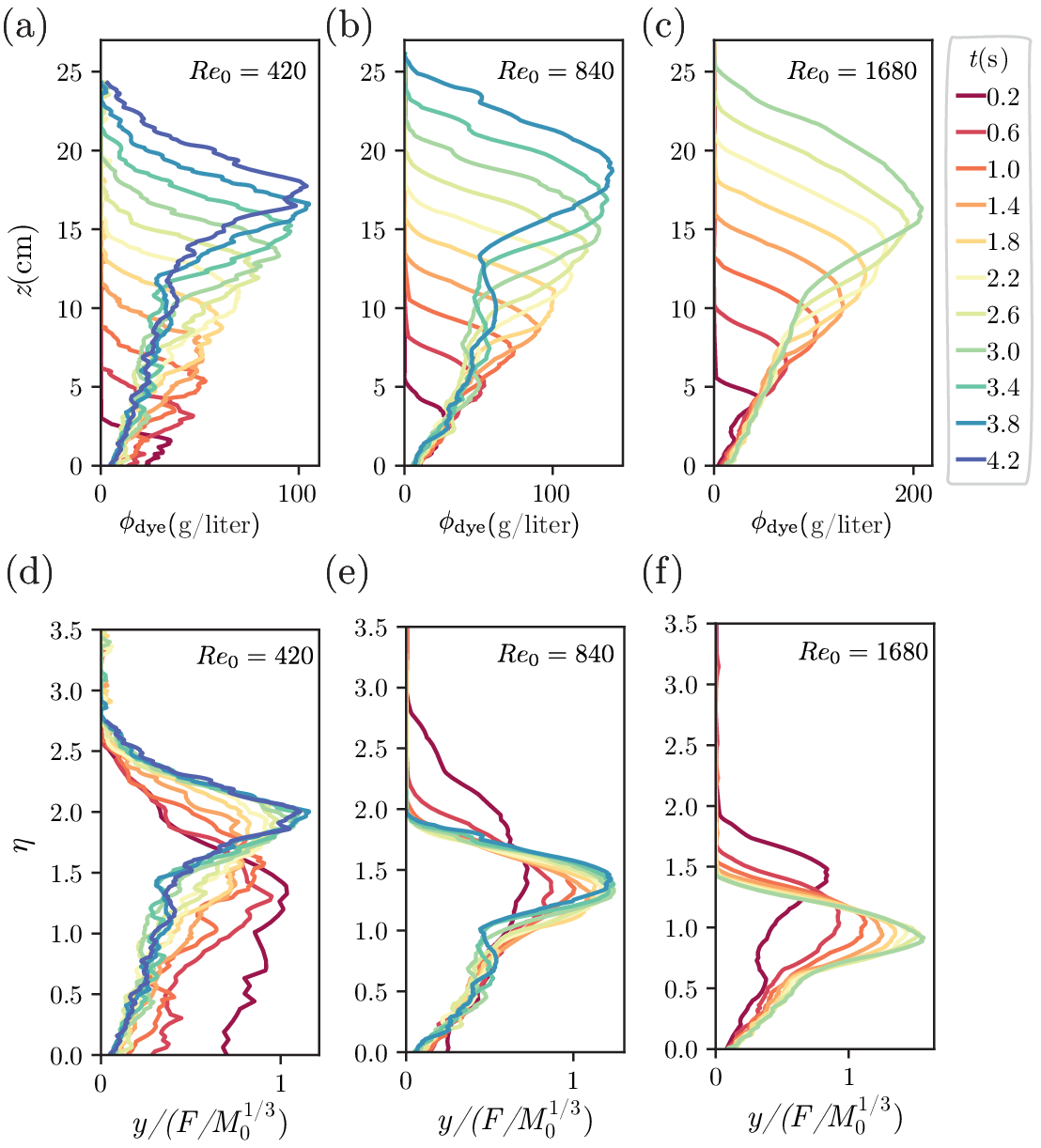}
\caption{The time evolution of the horizontally-integrated concentration profiles for the dye jets. $\phi_{\text{dye}}$ is the horizontally-integrated concentrations. (a--c) show the original profiles and (d--f) show the normalized ones from the corresponding case in (a--c) in terms of the similarity variables. In (d--f), $y=y(\eta)=\phi (z,t)t^{-1/3}$, $F$ is the source mass flux, and $M_0$ the initial momentum flux. All six figures share the same legend. } 
\label{img:cintjet_eth}
\end{figure}

\begin{figure}[h!bt]
\centering
\includegraphics[scale=1.2]{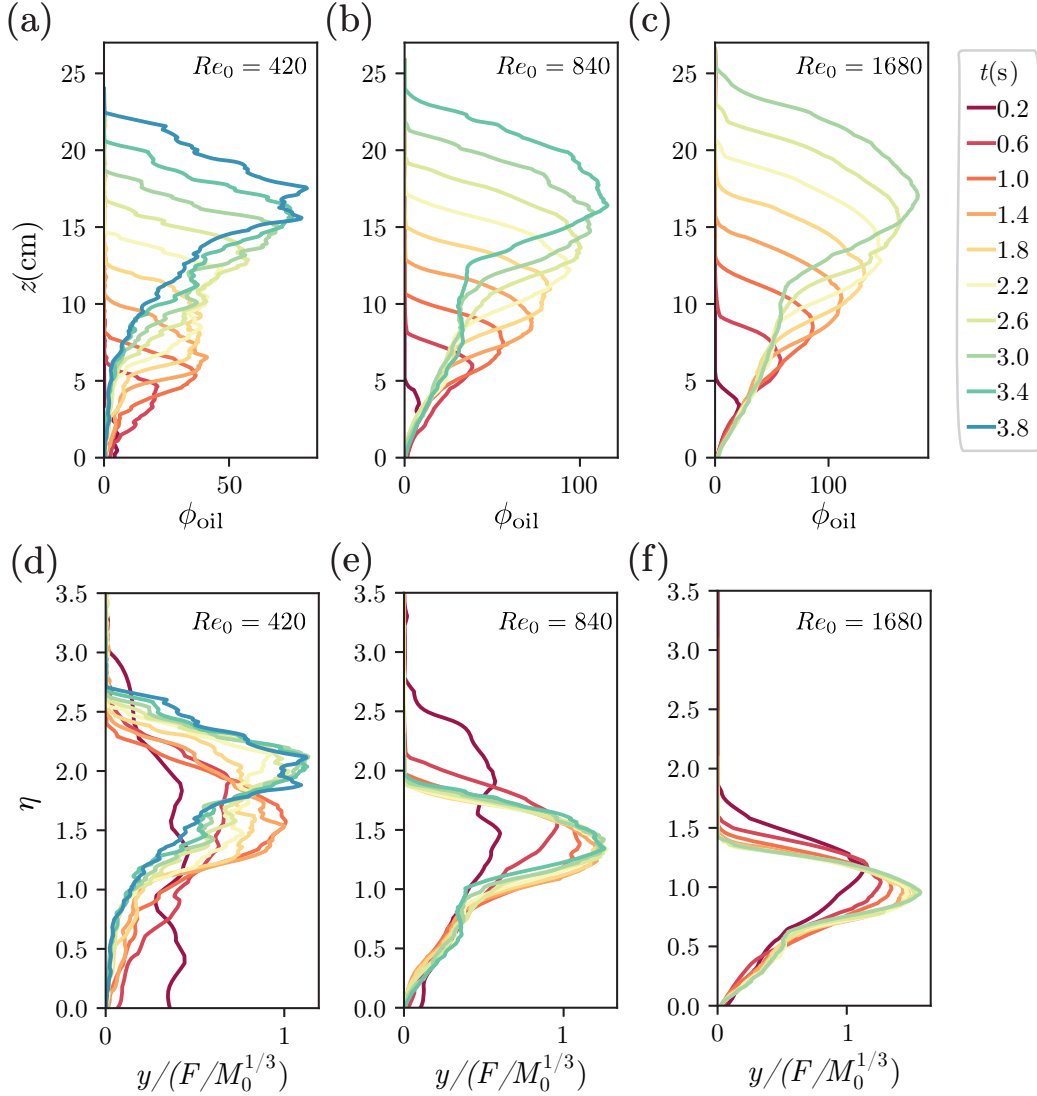}
\caption{The time evolution of the horizontally-integrated concentration profiles for the ouzo jets. $\phi_{\text{oil}}$ is the horizontally-integrated concentrations. (a--c) show the original profiles and (d--f) show the corresponding normalized ones in terms of the similarity variables. $F$ here is the flux of the summation of concentration, which is a function of time. All six figures share the same legend. } 
\label{img:cintjet_r100}
\end{figure}

\begin{figure}[h!bt]
\centering
\includegraphics[scale=1.05]{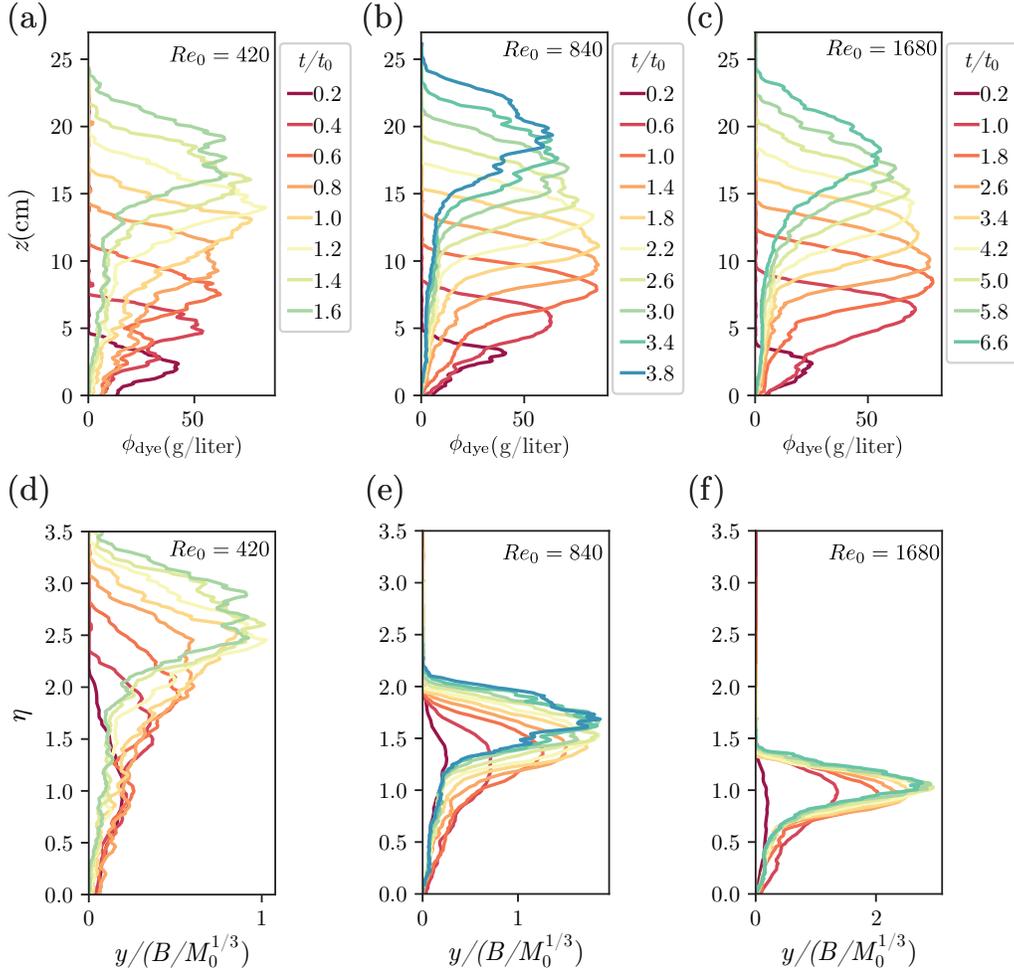}
\caption{The time evolution of the horizontally-integrated concentration profiles for the dye puffs. (a--c) show the original profiles and (d--f) show the normalized ones from the corresponding case in (a--c) in terms of the similarity variables. $y=y(\eta)=\phi (z,t)t^{-1/3}$, $M_0$ the initial momentum flux and $B$ is the summation of the concentrations. (a,d), (b,e), and (c,f) share the same legend respectively.} 
\label{img:cintpuff_eth}
\end{figure}

\begin{figure}[h!bt]
\centering
\includegraphics[scale=1.05]{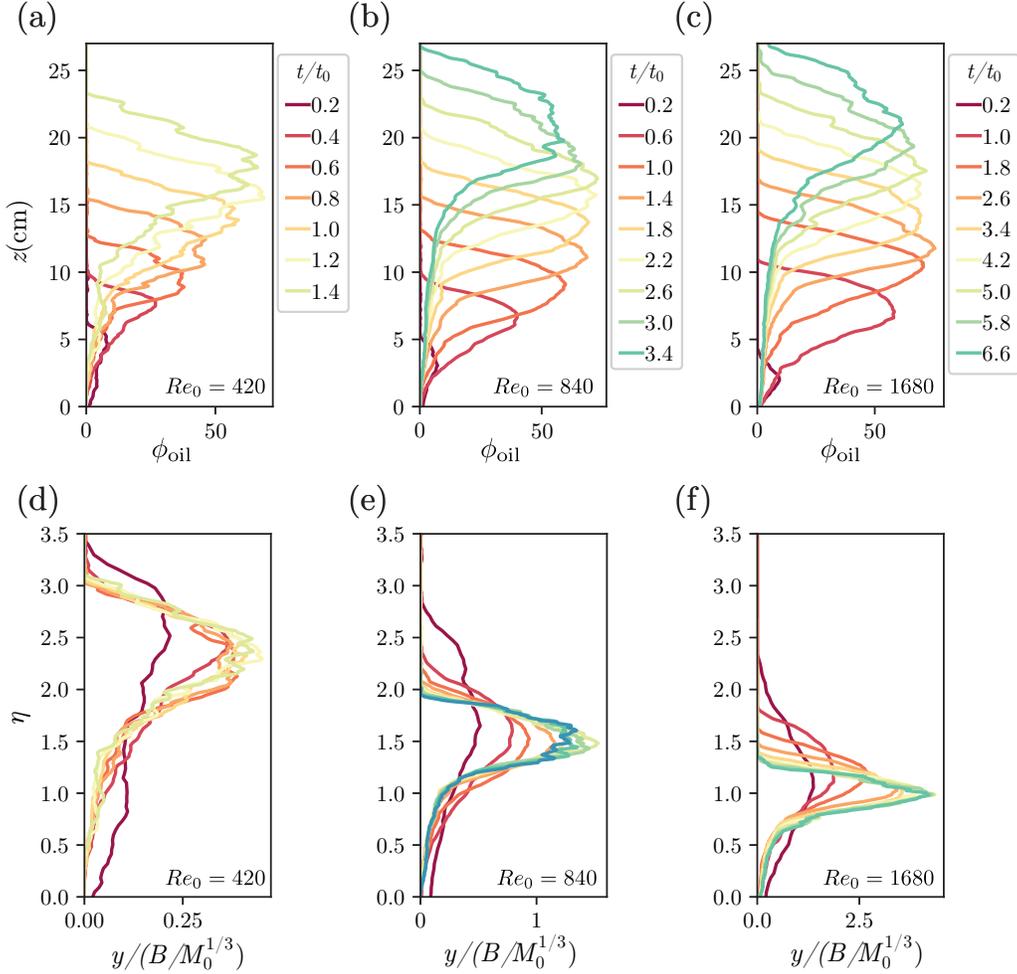}
\caption{The time evolution of the horizontally-integrated concentration profiles for the ouzo puffs. (a--c) show the original profiles and (d--f) show the normalized ones from the corresponding case in (a--c) in terms of the similarity variables. (a,d), (b,e), and (c,f) share the same legend respectively.} 
\label{img:cintpuff_r100}
\end{figure}

In the previous section, we approached entrainment and nucleation from a global perspective, considering the jet or the puff as a whole, and observing its evolution as a function of time. To investigate the temporal and spatial variation of the flows, we adapt the theoretical framework for quasi-2D jets and puffs in \citet{Landel2012b}, focusing on the streamwise evolution of the horizontally-integrated concentrations. In Figs.\ \ref {img:cintjet_eth}--\ref{img:cintpuff_r100} we show the experimental data for the streamwise distribution of the horizontally-integrated concentrations $\phi$ for the dye jets, the ouzo jets, the dye puffs, and the ouzo puffs, respectively. 

Figs.\ \ref {img:cintjet_eth}(a--c) present $\phi$ for the propagating dye jets, which consist of a vortex head and a trailing jet. The vortex head accumulates a large amount of $\phi$ due to its radial expansion, matching with the snapshots in Fig.\ \ref{img:snapjet}. It was shown \citep{Landel2012b} that the similarity variable $\eta$ can be formulated as $\eta = z/(t^{2/3}M_0^{1/3})$, suggesting that the front position $z$ scales with $t^{2/3}$. Although the local scaling exponent $n$ in Fig.\ \ref{img:zf}(c) is not precisely $2/3$, the normalized profiles in Figs.\ \ref {img:cintjet_eth}(d--f) show that the $2/3$ scaling works reasonably well for the propagation of $\phi$, at least for the two higher $Re_0$ cases. Note that the $F$ we use here is in fact the flux of $\iint_A C_{\text{dye}} \,dA$ instead of mass flux derived from Fig.\ \ref{img:csum}(a). The relation between the mass flux $\dot{m}_{\text{dye}}$ and the flux of $\iint_A C_{\text{dye}} \,dA$ can be expressed as:

\begin{align}
     \dot{m}_{\text{dye}} = P_1 \frac{d}{dt}\iint_A C_{\text{dye}} \,dA = P_1 F,
\label{eq:massfluxrelate}
\end{align}
where $P_1$ is a constant throughout the injection and $\dot{m}_{\text{dye}}$ and F can both be approximated as constant.

The profiles for the starting ouzo jets in Fig.\ \ref{img:cintjet_r100} look similar to the dye jets at first glance. The most noticeable difference is the more pronounced vortex head at the jet front. The continuous nucleation at the vortex head contributes to this feature of the ouzo jets. Also, in the early stage, the primary nucleation in the ouzo jets boosts the concentration levels, which is well captured by the change of profile from $t=0.2$s to $t=0.6$s in Figs.\ \ref {img:cintjet_r100}(a,d).

Without the trailing jet, the profiles between the dye puffs in Fig.\ \ref{img:cintpuff_eth} and the ouzo puffs in Fig.\ \ref{img:cintpuff_r100} become less differentiable. However, the concentration boost in the early stage is still pronounced and captured for all three cases, see the change of profile from $t/t_0=0.2$ to $t/t_0=0.4$, 0.6, and 1.0 in Figs.\ \ref{img:cintpuff_r100}(a--c), respectively.

Note that the $B$ we use here is the summation of concentration $\iint_A C_{\text{dye}} \,dA$ for dye and $\iint_A \widetilde{C}_{\text{oil}} \,dA$. The relation between the mass and the summation of concentrations for the dye case can be expressed as:

\begin{align}
 m_{\text{dye}} = P_2 \iint_A C_{\text{dye}} \,dA = P_2 B, 
\label{eq:massrelate}
\end{align}
where $P_2$ is constant throughout the injection.

%\clearpage
To model the experimental data presented above, we adapt the model presented in \citet{Landel2012b} for the constant flux case and the finite volume case,
\begin{align}
& \phi_{\text{jet}} = \frac{2Ft^{1/3}\eta^{1/2}}{3K_dM_0^{1/3}\Gamma\left[\frac{2}{3}(\frac{K_a}{K_d}+1)\right]}\Gamma\left[\frac{2}{3}\left(\frac{K_a}{K_d}-\frac{1}{2}\right),\frac{4\eta^{3/2}}{9K_d}\right], \label{eq:phijet}\\
& \phi_{\text{puff}} = \frac{2Bz^{1/2}}{3K_dM_0^{1/3}\Gamma\left[\frac{2}{3}(\frac{K_a}{K_d}+1)\right]}\Gamma\left[\frac{2}{3}\left(\frac{K_a}{K_d}-\frac{1}{2}\right),\frac{4z^{3/2}}{9K_d M_0^{1/2}t}\right] \notag\\ 
& -\Gamma\left[\frac{2}{3}\left(\frac{K_a}{K_d}-\frac{1}{2}\right),\frac{4z^{3/2}}{9K_d M_0^{1/2}(t-T_0)}\right]),\; T_0 < t \label{eq:phipuff}
\end{align}
where $\phi_{\text{jet}}$ is the horizontally-integrated concentration for the jet, $\phi_{\text{puff}}$ the horizontally-integrated concentration for the puff after the injection stops, $F$ is the flux of the concentration summation, $B$ the summation of the concentrations, $M_0$ the source momentum, $T_0$ the injection duration, $\eta = z/(t^{2/3}M_0^{1/3})$ the similarity variable, $K_a$ the advection coefficient, $K_d$ the dispersion coefficient, and $\Gamma$ the (incomplete) Gamma function,
\begin{align}
& \Gamma(s,x) = \int_x^{\infty} t^{s-1}e^{-t}dt, \\
& \Gamma(s) = \Gamma(s,0).
\end{align}
Note that a puff is essentially a starting jet during the injection.

As we mentioned earlier, our starting jets consist of a big vortex head and a trailing jet. Therefore, the model for a steady pure jet in Eq.\ \ref{eq:phijet} cannot accurately predict the integrated concentration in the vortex head. Knowing that the vortex head grows by its own entrainment and by the supply from the trailing jet \citep{Ai2006}, we combine Eqs.\ \ref{eq:phijet} and \ref{eq:phipuff}, aiming to better capture the profiles of a quasi-2D starting jet,

\begin{align}
 & \phi_{\text{sjet}} = a_j\phi_{\text{jet}} + a_p\phi_{\text{puff}}\\
 & a_j+a_p = 1
\label{eq:phisj}
\end{align}
where $a_j$ and $a_p$ are dimensionless fitting parameters and represent fractions of jet and puff. To determine $\phi_{\text{puff}}$, $B$ in Eq.\ \ref{eq:phipuff} is crucial. $B$ for a puff is simply the total injected mass (summation of concentration), which is not as straightforward for a jet. Here we choose the summation of concentration at the previous instant as the value substituted into Eq.\ \ref{eq:phipuff} for $B$. That is, $B(t) = B_{\text{exp}}(t-\Delta t)$, where $B_{\text{exp}}$ is the experimentally determined value in Eq.\ \ref{eq:massrelate} and $\Delta t$ is the time between frames..

At every instant, we perform optimization to minimize the difference between the experimental and the modelled results, obtaining the time-dependent advection coefficient $K_a$ and dispersion coefficient $K_d$. Although these coefficients should be constant based on the mathematical derivation \citep{Landel2012b}, the variation in time offers more accurate modelling of the dye case and the following attempt to estimate the amount of nucleation. 

\begin{figure}[h!bt]
\centering
\includegraphics[scale=1]{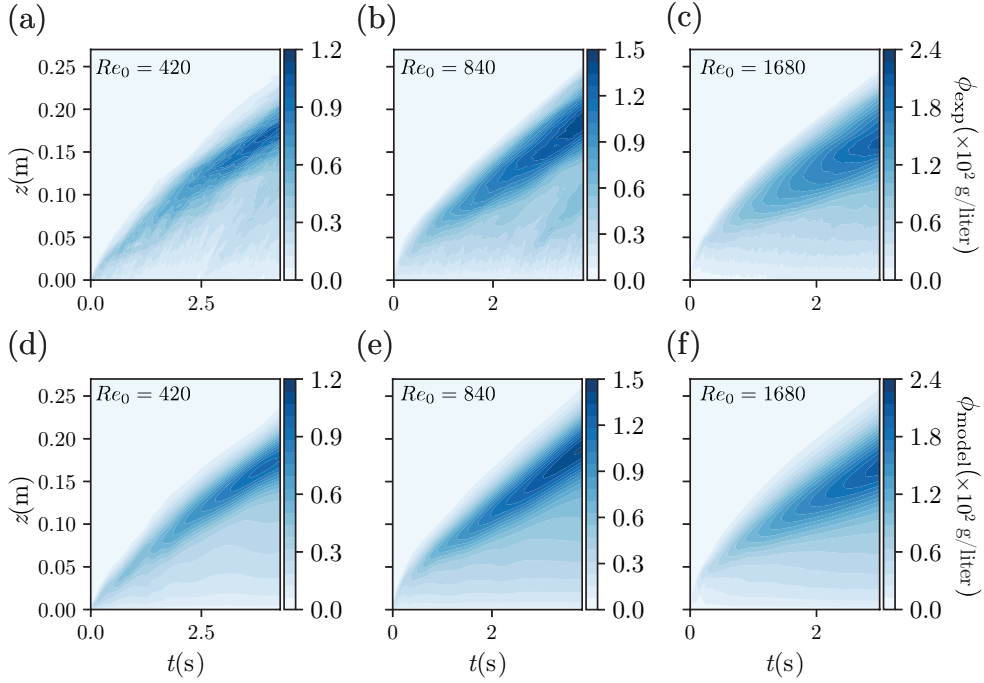}
\caption{The spatial-temporal contour plots for the horizontally-integrated concentrations for the dye jets. (a--c) show the results from the experiments, and (d--f) display the modelled results. The color code shows the horizontally-integrated concentrations $\phi$.} 
\label{img:phizt_jet_eth}
\end{figure}

\begin{figure}[h!bt]
\centering
\includegraphics[scale=1.15]{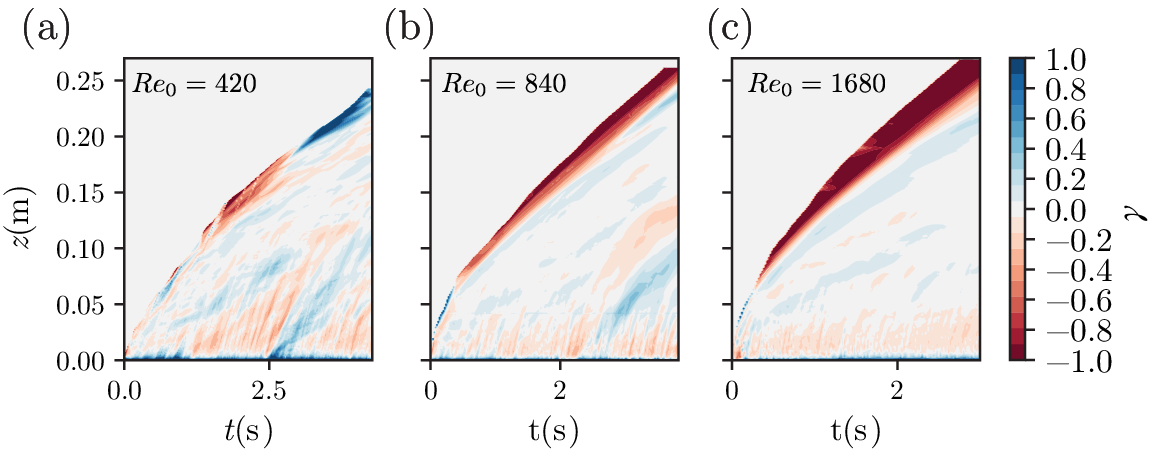}
\caption{The spatial-temporal contour plots for the deviation between the experiments and the models in Fig.\ \ref{img:phizt_jet_eth}. $\mathrm{\gamma} = (\phi_{\text{exp}} - \phi_{\text{model}})/\phi_{\text{model}}$ measures the degree of deviation. } 
\label{img:gamma_jet_eth}
\end{figure}

In Figs.\ \ref {img:phizt_jet_eth}(a--c) we construct the spatial-temporal contours of the experimentally determined $\phi_{\text{sjet}}$ using the profiles in Fig.\ \ref{img:cintjet_eth}. The contours for the modelled results are shown in Figs.\ \ref {img:phizt_jet_eth}(d--f), which show good agreement with the experimental results. We further present the deviation between the measurements and the model, plotting the spatial-temporal contours for the deviation factor $\gamma = (\phi_{\text{exp}}-\phi_{\text{model}})/\phi_{\text{model}}$ in Fig.\ \ref{img:gamma_jet_eth}. Small $\gamma$ in Fig.\ \ref{img:gamma_jet_eth} confirms that Eq.\ \ref{eq:phisj} accurately models the starting jets, except for the front area, where the model shows a more diffusive front and over-predicts $\phi$. 

\begin{figure}[h!bt]
\centering
\includegraphics[scale=1.1]{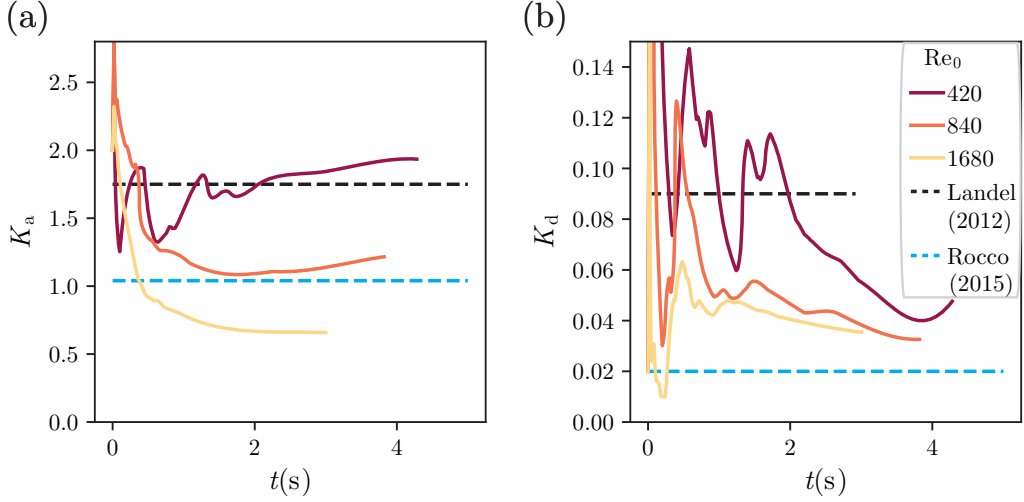}
\caption{The optimized advection coefficient $K_a$ and the dispersion coefficient $K_d$ for the starting dye jets. The black and the blue dashed lines indicate the results from \citet{Landel2012b, Rocco2015}} 
\label{img:k_jet_eth}
\end{figure}

Fig.\ \ref{img:k_jet_eth} displays the time-dependent fitted parameters for the modelled results in Figs.\ \ref {img:phizt_jet_eth}(d--f), namely $K_a$ and $K_d$. We also plot the fitted coefficients from \citet{Landel2012b, Rocco2015} for comparison. Note that \citet{Rocco2015} did not use the same framework we adapt from \citet{Landel2012b}. Both $K_a$ and $K_d$ do not align with the results reported by \citet{Landel2012b}, which we attribute to the difference between a steady pure jet and a starting buoyant jet. 

\begin{figure}[h!bt]
\centering
\includegraphics[scale=1.05]{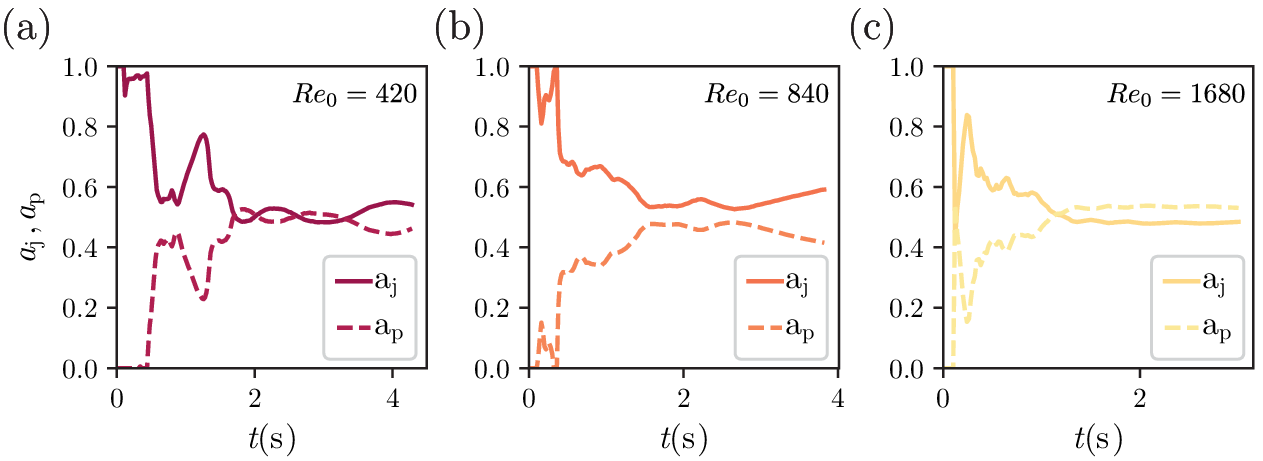}
\caption{The optimized jet and puff ratio, $a_j$ and $a_p$, for all three $Re_0$ for the dye starting jets.} 
\label{img:ajap_jet_eth}
\end{figure}

Fig.\ \ref{img:ajap_jet_eth} shows the evolution of the jet and puff ratio, $a_j$ and $a_p$ in Eq.\ \ref{eq:phisj}. While we cannot guarantee absolute physical relevance, these two parameters somewhat characterize the interaction between the vortex head (modelled by $a_p\phi_{\text{puff}}$) and the trailing jet (modelled by $a_j\phi_{\text{jet}}$). Fig.\ \ref{img:ajap_jet_eth} suggests that the vortex head quickly emerges after the injection, reaching around half of the total weight as the jet propagates.  

\begin{figure}[h!bt]
\centering
\includegraphics[scale=1.05]{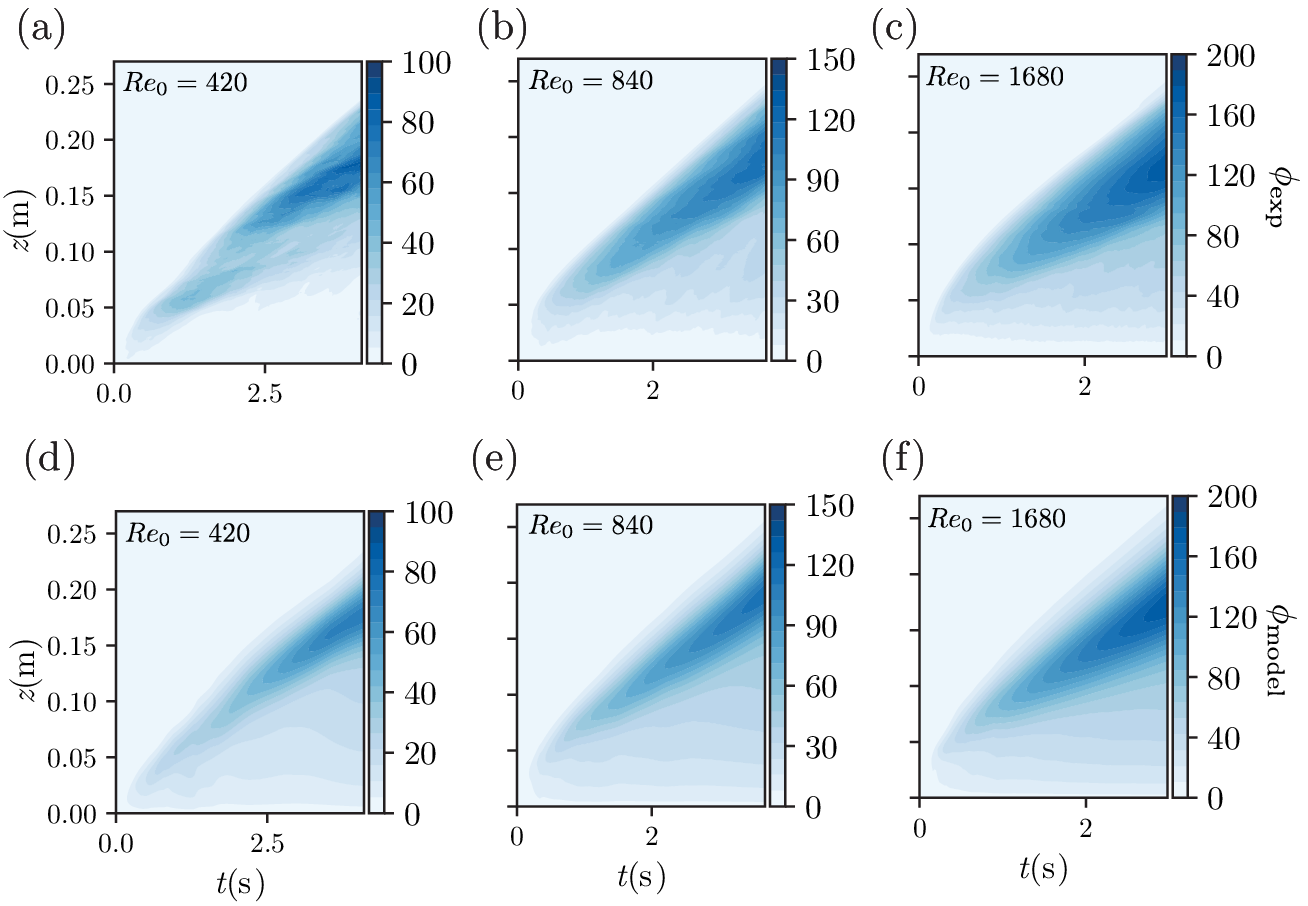}
\caption{The spatial-temporal contour plots for the horizontally-integrated concentrations for the ouzo jets, $\phi_{\text{oil}}$. (a--c) show the results from the experiments, and (d--f) display the modelled results.} 
\label{img:phizt_jet_r100}
\end{figure}

\begin{figure}[h!bt]
\centering
\includegraphics[scale=1.02]{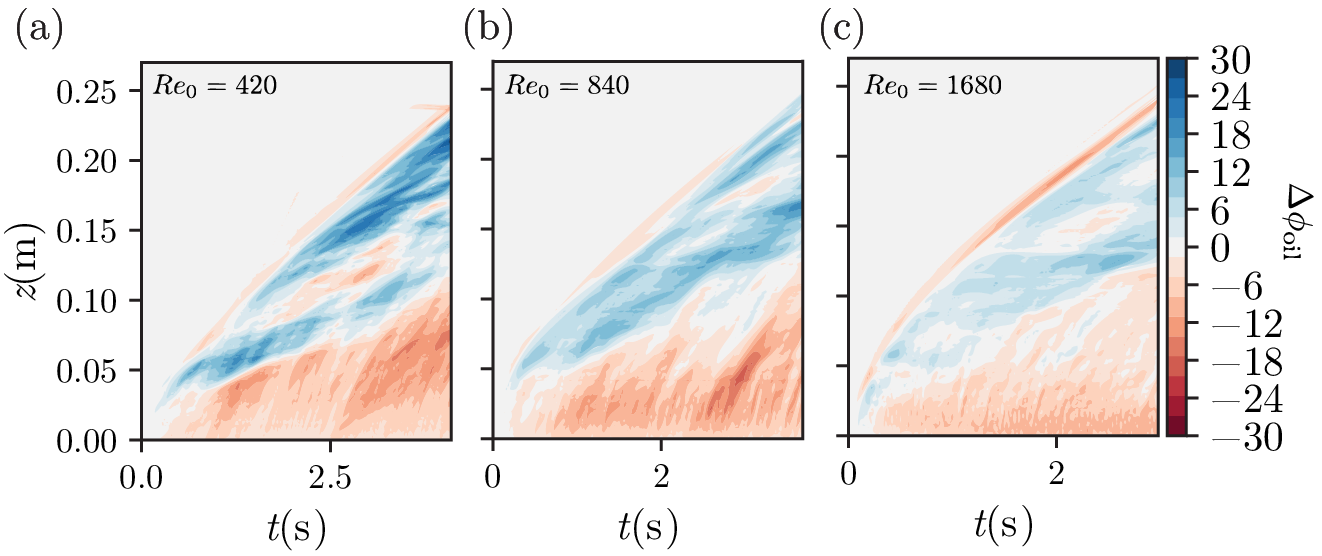}
\caption{The spatial-temporal contour plots for the amount of nucleation obtained from Fig.\ \ref{img:phizt_jet_r100}.} 
\label{img:nuc_jet_r100}
\end{figure}

To estimate the amount of nucleation for the ouzo jets, we assume that the ouzo jets and the dye jets are dynamically similar at the same $Re_0$. We constrain the fitting parameters $k_a$, $K_d$, $a_j$, and $a_p$ for the ouzo jets within $\pm{5}\%$ of the values for the dye jets, and perform the same optimization procedures. The difference between the experimental and the modelled results is then the amount of nucleation. While the mass flux $F$ is almost constant for the dye cases, it is monotonically increasing in time for the ouzo cases due to nucleation, see Figs.\ \ref{img:csum}(c) and \ref{img:cslope}(a). In order to estimate the amount of nucleation, the mass flux $F$ substituted into the model is determined by the corresponding experimental value of the previous instant, that is, $F(t)=\dot{m}_{\text{oil}}(t-\Delta t)$, where $\dot{m}_{\text{oil}}$ is determined in Fig.\ \ref{img:cslope}(a), and $\Delta t$ is the time between frames. The spatial-temporal contours for the experimental ouzo jets and the corresponding modelled ones are shown in Fig \ref{img:phizt_jet_r100}.

To account for the deviation factor $\gamma$ from the dye jets, we estimate the amount of nucleation by $\Delta \phi_{\text{oil}} = \phi_{\text{oil,exp}}-\gamma \phi_{\text{oil,model}}$. Fig .\ref{img:nuc_jet_r100} presents the spatial-temporal contours for the evolution of nucleation. Although we do not distinguish between the primary and the secondary nucleation in the model, the estimated nucleation here mostly comes from the secondary nucleation. The primary nucleation above the virtual origin serves as a mass influx, which is difficult to model in the early stage and thus only partially captured by the sharp color change slightly beyond $t=0$ in Fig .\ref{img:nuc_jet_r100}. We can identify that the increasing nucleation in time for the $Re_0=420$ case agrees with the monotonically increasing nucleation rate in Fig.\ \ref{img:cslope}(a). Moreover, the oil droplets nucleate along the entire vortex head and a small fraction of the trailing jet. The red areas at the tail are caused by the delayed appearance of the ouzo jets, see Fig.\ \ref{img:snapjet} for visualization. 

\begin{figure}[h!bt]
\centering
\includegraphics[scale=1.15]{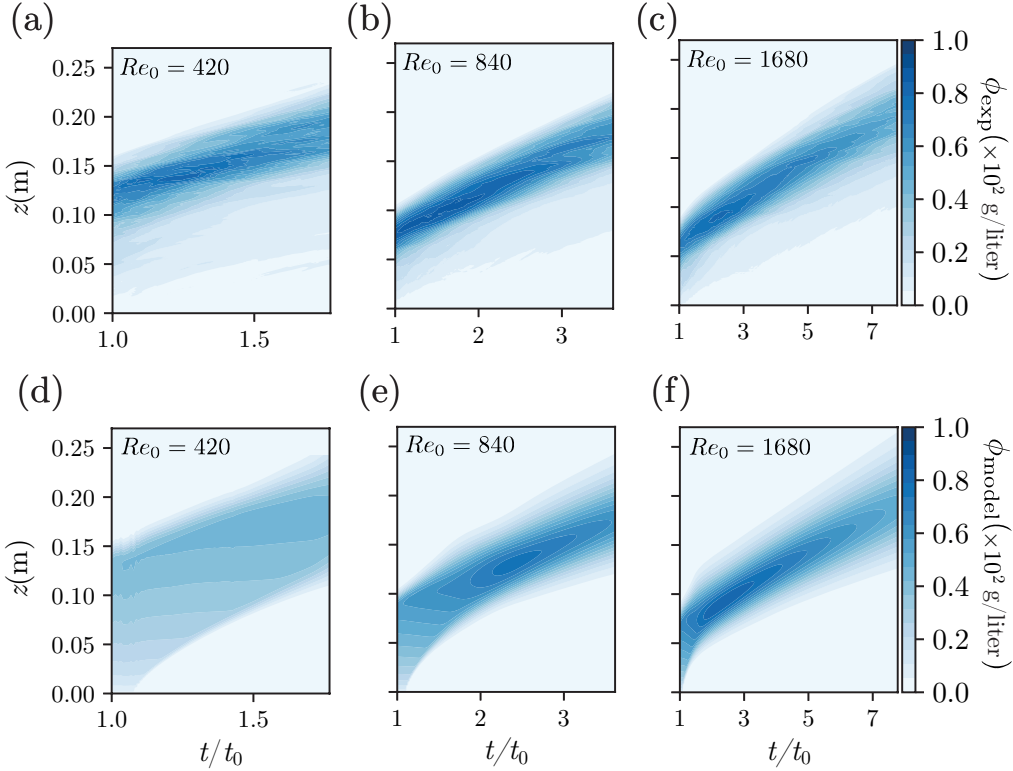}
\caption{The spatial-temporal contour plots for the horizontally-integrated concentrations for the dye puffs. (a--c) show the results from the experiments, and (d--f) display the modelled results.} 
\label{img:phizt_puff_eth}
\end{figure}

\begin{figure}[h!bt]
\centering
\includegraphics[scale=1.15]{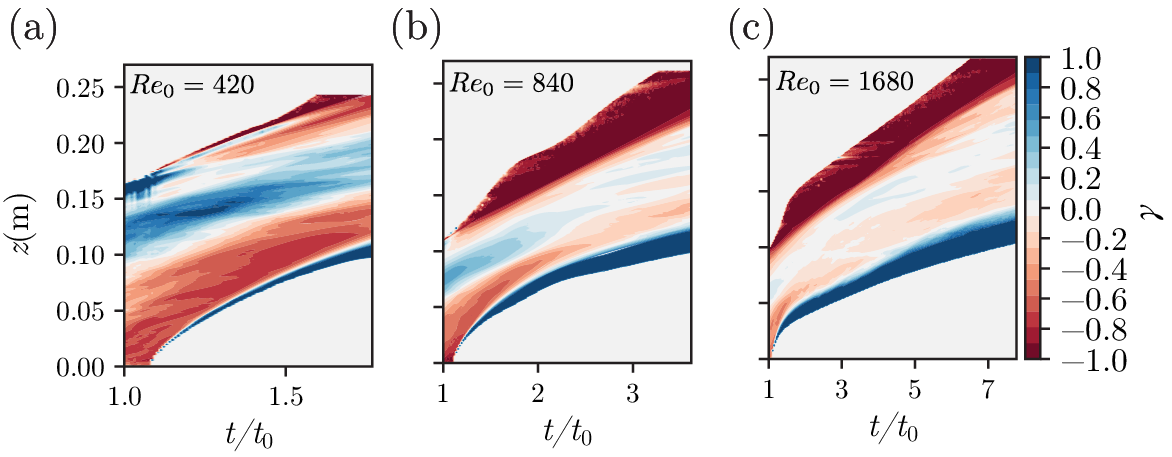}
\caption{The spatial-temporal contour plots for the deviation between the experiments and the models in Fig.\ \ref{img:phizt_puff_eth}. $\gamma = (\phi_{\text{exp}} - \phi_{\text{model}})/\phi_{\text{model}}$ measures the degree of deviation.} 
\label{img:gamma_puff_eth}
\end{figure}

For the dye puffs, we also construct the spatial-temporal contours in Figs.\ \ref {img:phizt_puff_eth}(a--c) using the experimental profiles presented in Fig.\ \ref{img:cintpuff_eth}. Applying Eq.\ \ref{eq:phijet} and \ref{eq:phipuff} to model the dye puffs, we obtained the modelled contours in Figs.\ \ref {img:phizt_puff_eth}(d--f). Calculating the deviation factor $\gamma$ between the experiments and the model, we obtain the spatial-temporal contours for the dye puffs shown in Fig.\ \ref{img:gamma_puff_eth}. Fig.\ \ref{img:gamma_puff_eth} shows that the model works better with increasing $Re_0$. Also, modelling the finite volume release with a rectangular source function \citep{Landel2012b}, Eq.\ \ref{eq:phipuff} takes more time to make the transition from a jet to a puff for $1\leq t/t_0 \leq 2$. For $t/t_0\geq 2$, the model agrees well with the experimental data, except for the diffusive front and tail.

\begin{figure}[h!bt]
\centering
\includegraphics[scale=1.1]{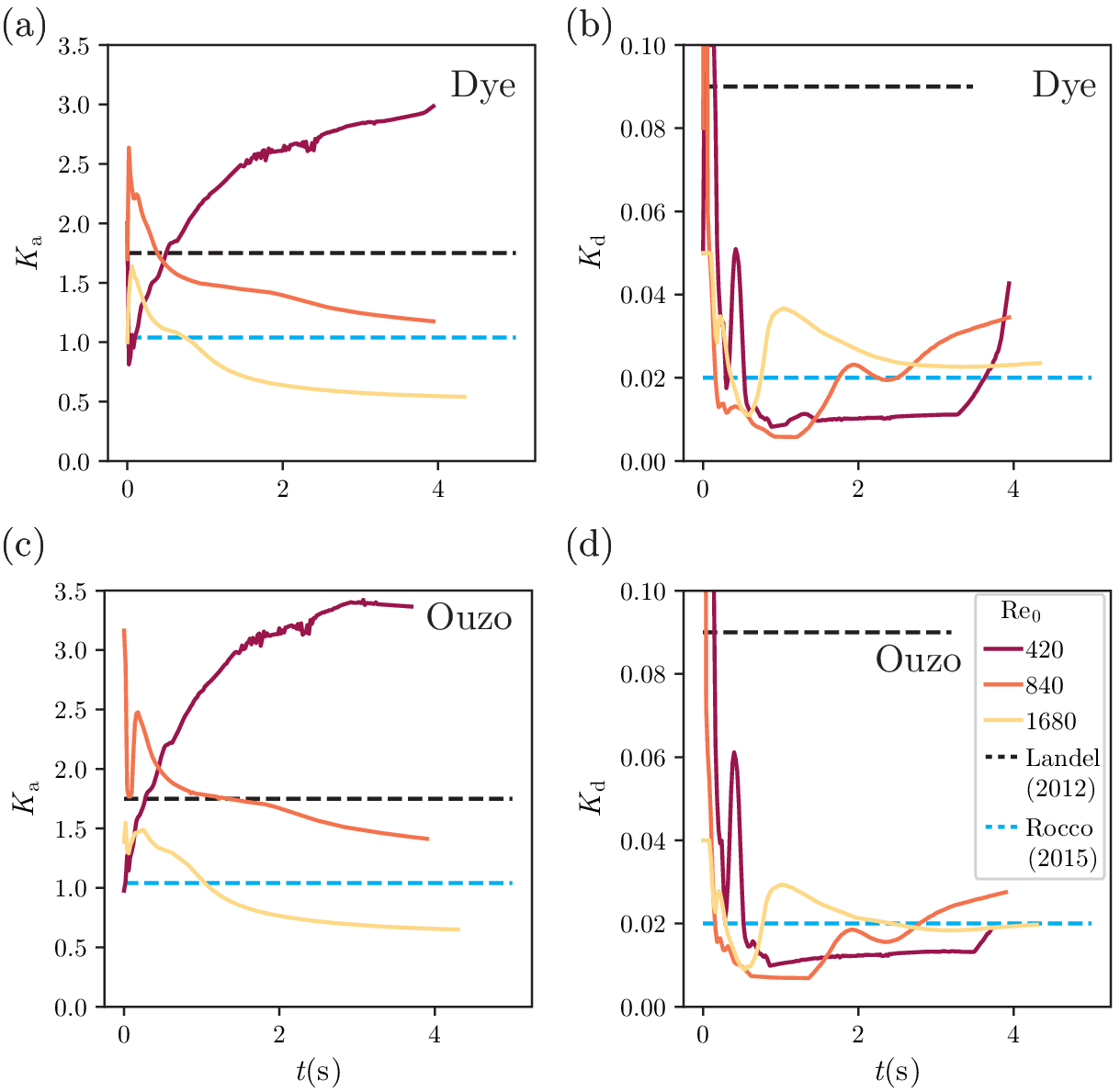}
\caption{The optimized advection coefficient $K_a$ and the dispersion coefficient $K_d$ for the dye puffs (a--b), and the ouzo puffs (c--d). The black and the blue dashed lines indicate the results from \citet{Landel2012b, Rocco2015}} 
\label{img:k_puff_both}
\end{figure}

\begin{figure}[h!bt]
\centering
\includegraphics[scale=1.15]{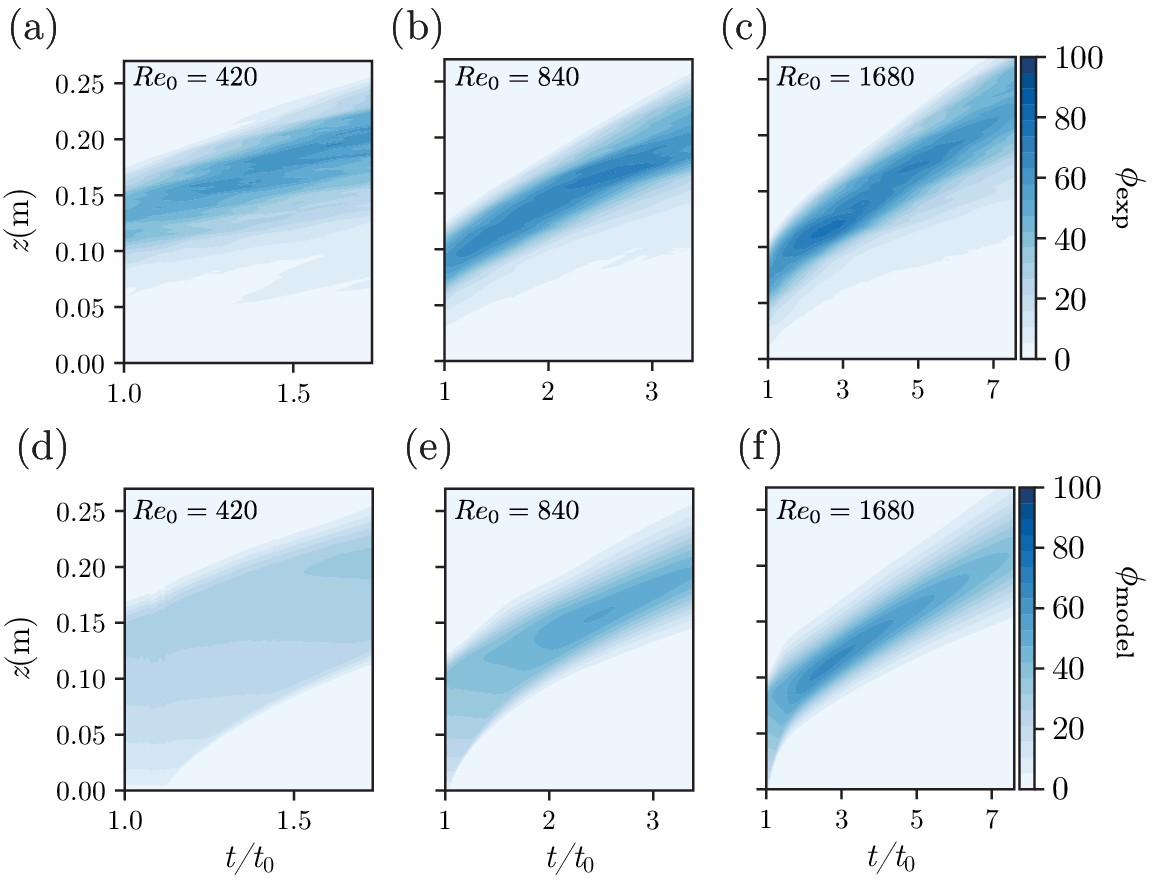}
\caption{The spatial-temporal contour plots for the horizontally-integrated concentrations for the ouzo puffs. (a--c) show the results from the experiments, and (d--f) display the modelled results.} 
\label{img:phizt_puff_r100}
\end{figure}

The fitted parameters $K_a$ and $K_d$ are then shown in Figs.\ \ref {img:k_puff_both}(a,b). While the values of the coefficients deviate from \citet{Landel2012b}, the slightly higher $K_a$ for puff than the corresponding $K_a$ for jet presented in Fig.\ \ref{img:k_jet_eth}(a) agrees with \citet{Landel2012b}. However, turbulent dispersion takes effect differently on the starting jets and on the puffs, which is different from previous findings \citep{Landel2012b}, see Figs.\ \ref{img:k_puff_both}(b) and \ref{img:k_jet_eth}(b). As shown in Figs.\ \ref{img:snapjet} and \ref{img:snappuff}, the highly buoyant quasi-2D puffs trigger the onset of the instability patterns quickly after injection, which might have a huge effect on the advection and dispersion of the flows as the Rayleigh Taylor instability greatly increases the length of the contour allowing for enhanced mixing. Therefore, it is not surprising to see that the highly buoyant quasi-2D starting jets and puffs behave differently than quasi-2D steady jets reported \citep{Landel2012b}.

Using the same approach we did to the starting jets, we attempted to estimate the amount of nucleation for the ouzo puffs using the $K_a$ and $K_d$ for the dye puffs. We first construct the spatial-temporal contours for the ouzo puffs in Figs.\ \ref {img:phizt_puff_r100}(a--c) from the profiles in Fig.\ \ref{img:cintpuff_r100}. However, when we imposed the $\pm{5}\%$ constraints on the fitting parameters, we found that the model puffs fall behind the measured ones, leading to a significant amount of estimated nucleation at the front, which we consider as an inaccurate estimation. To better capture the experimental data, we relax the constraints of the fitting parameters $K_a$ and $K_d$ to be $\pm{20}\%$ of the corresponding values obtained from the dye puffs, which better matches the propagation and the shape of the measured ouzo puffs. We present the fitted $K_a$ and $K_d$ in Figs.\ \ref {img:k_puff_both}(c,d), and the spatial-temporal contours for $\phi_{\text{model}}$ in Figs.\ \ref {img:phizt_puff_r100}(d--f). 

Comparing Figs.\ \ref {img:k_puff_both}(c,d) to Figs.\ \ref {img:k_puff_both}(a,b), the ouzo puffs have slightly stronger advection but slightly weaker dispersion. While the reason behind the weaker dispersion remains unclear, the stronger advection can be identified by Figs.\ \ref {img:cintpuff_eth}(a--c) and Figs.\ \ref {img:cintpuff_r100}(a--c), whose comparison shows faster advection of the puff front for the ouzo puffs. We attribute the enhanced front advection of $\phi_{\text{oil}}$ to the intense nucleation at the front upon mixing. For a dye puff, mixing at the front leads to the opposite effect on $\phi_{\text{dye}}$ due to dilution. Figs.\ \ref {img:frontcomp}(d--f) also show that the actual front for the dye puff and the corresponding ouzo puff propagates similarly. What makes them different in $\phi_{\text{oil}}$ and $K_a$ is the total enclosed concentration in the front area. 

\begin{figure}
\centering
\includegraphics[scale=1.15]{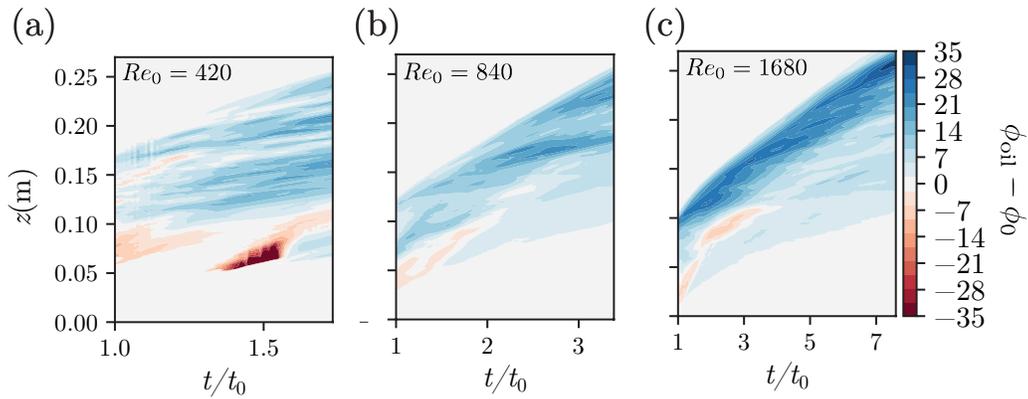}
\caption{The spatial-temporal contour plots for the accumulated nucleated mass obtained from Fig.\ \ref{img:phizt_puff_r100}. $\phi_{\text{oil}}-\phi_0$ is the accumulated nucleated mass after the injection stops.} 
\label{img:nuc_puff_r100}
\end{figure}

Since the puffs nucleate less than the jets due to weaker entrainment, the estimated instantaneous nucleation $\Delta \phi_{\text{oil}} = \phi_{\text{oil,exp}}-\gamma \phi_{\text{oil,model}}$ is sometimes at the same order of magnitude as and the deviation between the model and the experiment. Therefore, we display the accumulated additional nucleation after the injection stops, $\phi_{\text{oil}}-\phi_0$ in Fig.\ \ref{img:nuc_puff_r100}, where $\phi_{\text{oil}}$ is the accumulated nucleation, and $\phi_0$ is the amount of nucleation at $T_0$. To obtain $\phi_{\text{oil}}-\phi_0$, the summation of concentration $B$ in Eq.\ \ref{eq:phipuff} is the measured $m_{\text{oil}}$ at $t/t_0=1$ shown in Fig.\ \ref{img:csum}(d). Fig.\ \ref{img:nuc_puff_r100} shows that the droplets nucleate along the entire puff for all three $Re_0$, suggested by the appearance of dark patches as time evolves. Moreover, the $Re_0=1680$ case demonstrates enhanced nucleation at the puff front, which mainly occurs right after the injection stops. We attribute the enhanced nucleation for the $Re_0=1680$ case to the larger remaining momentum, which in turn leads to stronger entrainment and nucleation. 

\clearpage
\section{Conclusion}
We have experimentally investigated the propagation, entrainment, mixing, and subsequent nucleation in quasi-2D starting jets and puffs. We injected both dyed ethanol and an ouzo mixture to compare the case with and without nucleation. Using a light attenuation technique, we tracked the propagation of the jets and the puffs and measured the evolution of concentration. The quasi-2D starting jets and puffs go through various regimes, from the momentum-dominated jet and puff to the buoyancy-dominated plume and thermal. Due to a large density difference between the major component in the jet fluid and the ambient, namely ethanol and water, the jet front develops a pronounced head vortex, which later on becomes unstable due to a RT instability. 

Following the onset of turbulent entrainment, the ouzo jet flows only become visible above the virtual origin. This is where the primary nucleation occurs, which functions as the inlet of mass influx. Beyond the primary nucleation site, continuous entrainment leads to secondary nucleation and dilution at the same time, which act against each other in determining concentration. Considering a jet or a puff as a whole, the time evolution of the nucleation rate closely follows that of the entrained volume flux, indicating the dominant role of turbulent entrainment in inducing solvent exchange. 

Adapting the analytical framework developed by \citet{Landel2012b}, we successfully model the horizontally-integrated dye concentration measured from the experiments, obtaining the fitted advection and dispersion coefficients, $K_a$ \& $K_d$, for both the starting jets and the puffs, and the jet and puff ratio, $a_j$ \& $a_p$, for the starting jets. These fitted parameters from the dye cases are then plugged into the corresponding ouzo case with $\pm{5}\%$ -- $\pm{20}\%$ relaxation. Combining the fitted parameters with a careful selection of the mass flux $F$ and the total mass $B$ for different scenarios, we estimate the spatial-temporal evolution of the nucleation rate for the jets, and that of the accumulated nucleated mass for the puffs. The estimations not only agree with the temporal evolution presented in Fig.\ \ref{img:cslope}, but also reveals the spatial distribution of nucleation for various scenarios, from the effect of $Re_0$ to the continuous or finite injection. The presented framework takes our understanding of solvent exchange in turbulent flows to the next level.
%---------------------------
\section*{Acknowledgement}
The authors acknowledge the funding by ERC Advanced Grant Diffusive Droplet Dynamics (DDD) with Project No.740479 and Netherlands Organisation for Scientific Research (NWO) through the Multiscale Catalytic Energy Conversion (MCEC) research center.

We thank G.W. Bruggert, M. Bos, and T. Zijlstra for their technical support in building the setup. 
%---------------------------
\section*{Appendix A: Calibration curves}
\begin{figure}[h!bt]
\centering
\includegraphics[scale=1.1]{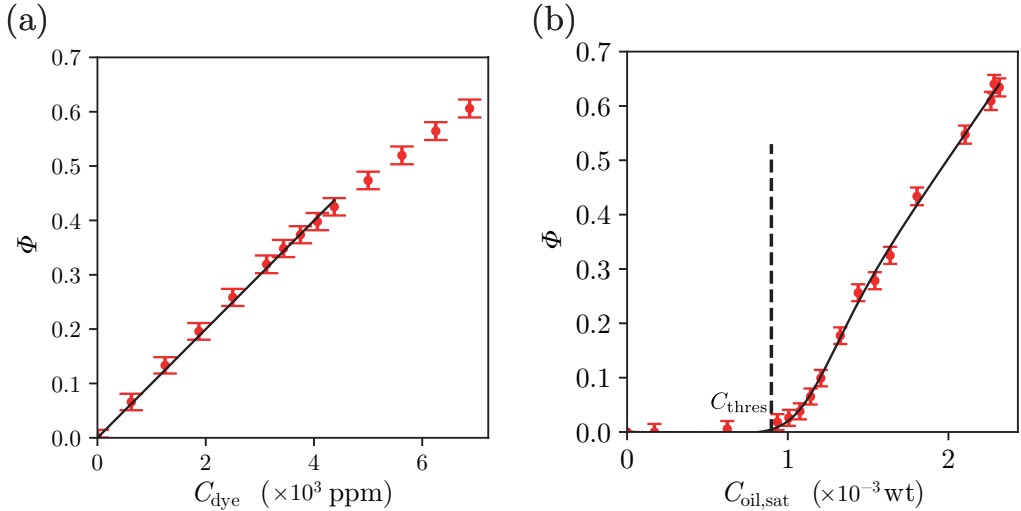}
\caption{The calibration curve for (a) the dyed ethanol and (b) the ouzo mixture. $\Phi=\log(I_{\text{ref}}/I)$ is the degree of light attenuation. The red points are the measured data from the calibration, the black curves are the fitted calibration curves, and the error bars represent the standard deviation of $\Phi$. The abscissa $C_{\text{oil,sat}}$ in (b) is the oversaturation of the oil. $C_{\text{thres}}$ denotes the threshold oversaturation used to normalized oil oversaturation.} 
\label{img:calicurve3}
\end{figure}
In \S2.2 we mentioned that we have used in-situ calibration curves to convert the recorded light attenuation to dye concentration or oversaturation of the nucleated oil. For the dye case, we obtained calibration data points shown in Fig.\ \ref{img:calicurve3}(a), which can be fitted with a linear curve for concentration below \SI{4000}ppm. For the ouzo case, the calibration data points are obtained using the phase diagram discussed in \S2.2 and Fig.\ \ref{img:phasegram}, which we fit by the empirical function,
\begin{equation}
    \Phi(C) = \log \left( I_{\text{ref}}/I \right)= \frac{a_0}{1+a_1e^{a_2(C-a_3)}}+\frac{a_4(C-a_5)}{1+a_1e^{a_2(C-a_3)}},
\label{eq:LA3}
\end{equation}
where $\Phi$ is the light attenuation level, $C$ the oil concentration (or oversaturation), and $a_1$--$a_5$ the fitted coefficients.
A calibration curve for the ouzo mixture is shown in Fig.\ \ref{img:calicurve3}(b). We define $C_{\text{thres}}$ to be the point where $\Phi$ reaches 1\% of its peak. Note that the calibration curve shown here is a local calibration curve for one image unit of 5 $\times$ 5 pixels. We have obtained in total 204 $\times$ 204 local calibration curves within the recording domain.
%---------------------------
\section*{Appendix B: Front propagation of the ouzo cases}
\begin{figure}[h!bt]
\centering
\includegraphics[scale=1.15]{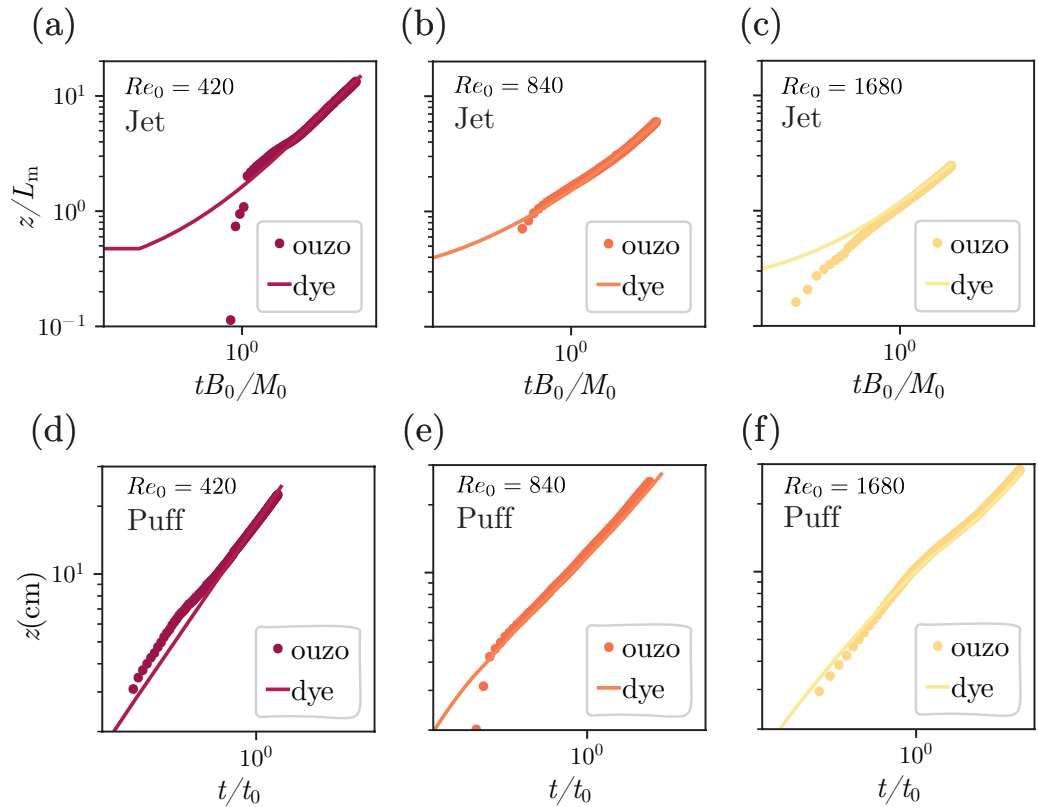}
\caption{The comparison of the front propagation between the ouzo case and the dye case under various scenarios.} 
\label{img:frontcomp}
\end{figure}
We argued in \S3.2 that the dyed ethanol and the ouzo mixture are similar in dynamics, and used the front propagation of the dyed ethanol to represent that of the ouzo mixture. However, we found in \S3.4 that the ouzo puff has a larger advection coefficient $K_a$ than the corresponding dye puff, which seems contradictory to our argument. To validate our argument, we show the comparison of the front propagation between the ouzo case and the dye case for each $Re_0$, and for jet and puff respectively in Fig.\ \ref{img:frontcomp}. 

Fig.\ \ref{img:frontcomp} indicates that the front trajectories of the two jet fluids are almost identical for all the scenarios, except for the deviation in the initial period. The deviation can be attributed to the delayed appearance of the ouzo jets shown in Figs.\ \ref {img:snapjet} and \ref{img:snappuff}. The initial leading front trace of the ouzo puff in Fig.\ \ref{img:frontcomp}(d), however, remains unclear, which does not compromise our results in this chapter. 

\bibliographystyle{elsarticle-harv} 
\bibliography{main}

\begin{thebibliography}{40}
\expandafter\ifx\csname natexlab\endcsname\relax\def\natexlab#1{#1}\fi
\providecommand{\url}[1]{\texttt{#1}}
\providecommand{\href}[2]{#2}
\providecommand{\path}[1]{#1}
\providecommand{\DOIprefix}{doi:}
\providecommand{\ArXivprefix}{arXiv:}
\providecommand{\URLprefix}{URL: }
\providecommand{\Pubmedprefix}{pmid:}
\providecommand{\doi}[1]{\href{http://dx.doi.org/#1}{\path{#1}}}
\providecommand{\Pubmed}[1]{\href{pmid:#1}{\path{#1}}}
\providecommand{\bibinfo}[2]{#2}
\ifx\xfnm\relax \def\xfnm[#1]{\unskip,\space#1}\fi
%Type = Article
\bibitem[{Ai et~al.(2006)Ai, Law and Yu}]{Ai2006}
\bibinfo{author}{Ai, J.}, \bibinfo{author}{Law, A.}, \bibinfo{author}{Yu, S.},
  \bibinfo{year}{2006}.
\newblock \bibinfo{title}{On {B}oussinesq and non-{B}oussinesq starting forced
  plumes}.
\newblock \bibinfo{journal}{J. Fluid Mech.} \bibinfo{volume}{558},
  \bibinfo{pages}{357--386}.
%Type = Article
\bibitem[{Ai et~al.(2005)Ai, Yu, Law, K. and Chua}]{Ai2005}
\bibinfo{author}{Ai, J.J.}, \bibinfo{author}{Yu, S.C.M.},
  \bibinfo{author}{Law}, \bibinfo{author}{K., A.W.}, \bibinfo{author}{Chua,
  L.P.}, \bibinfo{year}{2005}.
\newblock \bibinfo{title}{Vortex dynamics in starting square water jets}.
\newblock \bibinfo{journal}{Phys. Fluids} \bibinfo{volume}{17},
  \bibinfo{pages}{014106}.
%Type = Article
\bibitem[{Balachandar et~al.(1999)Balachandar, Tachie and
  Chu}]{Balachandar1999}
\bibinfo{author}{Balachandar, R.}, \bibinfo{author}{Tachie, M.F.},
  \bibinfo{author}{Chu, V.H.}, \bibinfo{year}{1999}.
\newblock \bibinfo{title}{Concentration profiles in shallow turbulent wakes}.
\newblock \bibinfo{journal}{J. Fluids Eng.} \bibinfo{volume}{121},
  \bibinfo{pages}{34--43}.
%Type = Article
\bibitem[{Bourouiba(2021)}]{Bourouiba2021}
\bibinfo{author}{Bourouiba, L.}, \bibinfo{year}{2021}.
\newblock \bibinfo{title}{The fluid dynamics of disease transmission}.
\newblock \bibinfo{journal}{Annu. Rev. Fluid Mech.} \bibinfo{volume}{53},
  \bibinfo{pages}{473--508}.
%Type = Article
\bibitem[{Chen and Jirka(1995)}]{Chen1995}
\bibinfo{author}{Chen, D.}, \bibinfo{author}{Jirka, G.H.},
  \bibinfo{year}{1995}.
\newblock \bibinfo{title}{Experimental study of plane turbulent wakes in a
  shallow water layer}.
\newblock \bibinfo{journal}{Fluid Dyn. Res.} \bibinfo{volume}{16},
  \bibinfo{pages}{11}.
%Type = Article
\bibitem[{Chong et~al.(2021)Chong, Ng, Hori, Yang, Verzicco and
  Lohse}]{Chong2021}
\bibinfo{author}{Chong, K.L.}, \bibinfo{author}{Ng, C.S.},
  \bibinfo{author}{Hori, N.}, \bibinfo{author}{Yang, R.},
  \bibinfo{author}{Verzicco, R.}, \bibinfo{author}{Lohse, D.},
  \bibinfo{year}{2021}.
\newblock \bibinfo{title}{Extended lifetime of respiratory droplets in a
  turbulent vapor puff and its implications on airborne disease transmission}.
\newblock \bibinfo{journal}{Phys. Rev. Lett.} \bibinfo{volume}{126},
  \bibinfo{pages}{034502}.
%Type = Article
\bibitem[{Cossali et~al.(2001)Cossali, Coghe and Araneo}]{Cossali2001}
\bibinfo{author}{Cossali, G.E.}, \bibinfo{author}{Coghe, A.},
  \bibinfo{author}{Araneo, L.}, \bibinfo{year}{2001}.
\newblock \bibinfo{title}{Near-field entrainment in an impulsively started
  turbulent gas jet}.
\newblock \bibinfo{journal}{AIAA J.} \bibinfo{volume}{39},
  \bibinfo{pages}{1113--1122}.
%Type = Article
\bibitem[{Diez et~al.(2003)Diez, Bernal and Faeth}]{Diez2003}
\bibinfo{author}{Diez, F.J.}, \bibinfo{author}{Bernal, L.P.},
  \bibinfo{author}{Faeth, G.M.}, \bibinfo{year}{2003}.
\newblock \bibinfo{title}{Round turbulent thermals, puffs, starting plumes and
  starting jets in uniform crossflow}.
\newblock \bibinfo{journal}{J. Heat Transfer.} \bibinfo{volume}{125},
  \bibinfo{pages}{1046--1057}.
%Type = Article
\bibitem[{Dracos et~al.(1992)Dracos, Giger and Jirka}]{Dracos1992}
\bibinfo{author}{Dracos, T.}, \bibinfo{author}{Giger, M.},
  \bibinfo{author}{Jirka, G.H.}, \bibinfo{year}{1992}.
\newblock \bibinfo{title}{Plane turbulent jets in a bounded fluid layer}.
\newblock \bibinfo{journal}{J. Fluid Mech.} \bibinfo{volume}{241},
  \bibinfo{pages}{587--614}.
%Type = Article
\bibitem[{Fischer(1976)}]{Fischer1976}
\bibinfo{author}{Fischer, H.}, \bibinfo{year}{1976}.
\newblock \bibinfo{title}{Mixing and dispersion in estuaries}.
\newblock \bibinfo{journal}{Annu. Rev. Fluid Mech.} \bibinfo{volume}{8},
  \bibinfo{pages}{107--133}.
%Type = Article
\bibitem[{Fischer et~al.(1979)Fischer, List, Koh, Imberger and
  Brooks}]{Fischer1979}
\bibinfo{author}{Fischer, H.B.}, \bibinfo{author}{List, E.J.},
  \bibinfo{author}{Koh, R.C.Y.}, \bibinfo{author}{Imberger, J.},
  \bibinfo{author}{Brooks, N.H.}, \bibinfo{year}{1979}.
\newblock \bibinfo{title}{Mixing in inland and coastal waters}.
\newblock \bibinfo{journal}{Academic} .
%Type = Article
\bibitem[{Ghaem-Maghami and Johari(2007)}]{Ghaem-Maghami2007}
\bibinfo{author}{Ghaem-Maghami, E.}, \bibinfo{author}{Johari, H.},
  \bibinfo{year}{2007}.
\newblock \bibinfo{title}{Concentration field measurements within isolated
  turbulent puffs}.
\newblock \bibinfo{journal}{J. Fluids Eng.} \bibinfo{volume}{129},
  \bibinfo{pages}{194--199}.
%Type = Article
\bibitem[{Giger et~al.(1991)Giger, Dracos and Jirka}]{Giger1991}
\bibinfo{author}{Giger, M.}, \bibinfo{author}{Dracos, T.},
  \bibinfo{author}{Jirka, G.H.}, \bibinfo{year}{1991}.
\newblock \bibinfo{title}{Entrainment and mixing in plane turbulent jets in
  shallow water}.
\newblock \bibinfo{journal}{J. Hydraul. Res.} \bibinfo{volume}{29},
  \bibinfo{pages}{615--642}.
%Type = Article
\bibitem[{Glezer(1988)}]{Glezer1988}
\bibinfo{author}{Glezer, A.}, \bibinfo{year}{1988}.
\newblock \bibinfo{title}{The formation of vortex rings}.
\newblock \bibinfo{journal}{Phys. Fluids} \bibinfo{volume}{31},
  \bibinfo{pages}{3532--3542}.
%Type = Article
\bibitem[{Gutmark and Wygnanski(1976)}]{Gutmark1976}
\bibinfo{author}{Gutmark, E.}, \bibinfo{author}{Wygnanski, I.},
  \bibinfo{year}{1976}.
\newblock \bibinfo{title}{The planar turbulent jet}.
\newblock \bibinfo{journal}{J. Fluid Mech.} \bibinfo{volume}{73},
  \bibinfo{pages}{465--495}.
%Type = Article
\bibitem[{Hajian and Hardt(2015)}]{Hajian2015}
\bibinfo{author}{Hajian, R.}, \bibinfo{author}{Hardt, S.},
  \bibinfo{year}{2015}.
\newblock \bibinfo{title}{Formation and lateral migration of nanodroplets via
  solvent shifting in a microfluidic device}.
\newblock \bibinfo{journal}{Microfluid Nanofluid} \bibinfo{volume}{19},
  \bibinfo{pages}{1281--1296}.
%Type = Article
\bibitem[{Hannah(2017)}]{Hannah2017}
\bibinfo{author}{Hannah, W.M.}, \bibinfo{year}{2017}.
\newblock \bibinfo{title}{Entrainment versus dilution in tropical deep
  convection}.
\newblock \bibinfo{journal}{J. Atmos. Sci.} \bibinfo{volume}{74},
  \bibinfo{pages}{3725--3747}.
%Type = Article
\bibitem[{Hassanzadeh et~al.(2021)Hassanzadeh, Eslami and
  Taghavi}]{Hassanzadeh2021}
\bibinfo{author}{Hassanzadeh, H.}, \bibinfo{author}{Eslami, A.},
  \bibinfo{author}{Taghavi, S.M.}, \bibinfo{year}{2021}.
\newblock \bibinfo{title}{Positively buoyant jets: Semiturbulent to fully
  turbulent regimes}.
\newblock \bibinfo{journal}{Phys. Rev. Fluids} \bibinfo{volume}{6},
  \bibinfo{pages}{054501}.
%Type = Article
\bibitem[{Jirka(2001)}]{Jirka2001}
\bibinfo{author}{Jirka, G.H.}, \bibinfo{year}{2001}.
\newblock \bibinfo{title}{Large scale flow structures and mixing processes in
  shallow flows}.
\newblock \bibinfo{journal}{J. Hydraul. Res.} \bibinfo{volume}{39},
  \bibinfo{pages}{567--573}.
%Type = Article
\bibitem[{Johari and Motelvalli(1993)}]{Johari1993}
\bibinfo{author}{Johari, H.}, \bibinfo{author}{Motelvalli, V.},
  \bibinfo{year}{1993}.
\newblock \bibinfo{title}{Flame length measurements of burning fuel puffs}.
\newblock \bibinfo{journal}{Combust. Sci. Technol.} \bibinfo{volume}{94},
  \bibinfo{pages}{229--244}.
%Type = Article
\bibitem[{Landel et~al.(2012a)Landel, Caulfield and Woods}]{Landel2012a}
\bibinfo{author}{Landel, J.}, \bibinfo{author}{Caulfield, C.},
  \bibinfo{author}{Woods, A.}, \bibinfo{year}{2012}a.
\newblock \bibinfo{title}{Meandering due to large eddies and the statistically
  self-similar dynamics of quasi-two-dimensional jets}.
\newblock \bibinfo{journal}{J. Fluid Mech.} \bibinfo{volume}{692},
  \bibinfo{pages}{347--368}.
%Type = Article
\bibitem[{Landel et~al.(2012b)Landel, Caulfield and Woods}]{Landel2012b}
\bibinfo{author}{Landel, J.}, \bibinfo{author}{Caulfield, C.},
  \bibinfo{author}{Woods, A.}, \bibinfo{year}{2012}b.
\newblock \bibinfo{title}{Streamwise dispersion and mixing in
  quasi-two-dimensional steady turbulent jets}.
\newblock \bibinfo{journal}{J. Fluid Mech.} \bibinfo{volume}{771},
  \bibinfo{pages}{212--258}.
%Type = Article
\bibitem[{Lee et~al.(2022)Lee, Sun and Lohse}]{Lee2022}
\bibinfo{author}{Lee, Y.}, \bibinfo{author}{Sun, C.and~Huisman, S.},
  \bibinfo{author}{Lohse, D.}, \bibinfo{year}{2022}.
\newblock \bibinfo{title}{Micro-droplet nucleation through solvent exchange in
  a turbulent buoyant jet}.
\newblock \bibinfo{journal}{J. Fluid Mech.} \bibinfo{volume}{943},
  \bibinfo{pages}{A11}.
%Type = Article
\bibitem[{Li et~al.(2021)Li, Chong, Bazyar, Lammertink and Lohse}]{Li2021}
\bibinfo{author}{Li, Y.}, \bibinfo{author}{Chong, K.}, \bibinfo{author}{Bazyar,
  H.}, \bibinfo{author}{Lammertink, R.}, \bibinfo{author}{Lohse, D.},
  \bibinfo{year}{2021}.
\newblock \bibinfo{title}{Universality in microdroplet nucleation during
  solvent exchange in {{Hele-Shaw}}-like channels}.
\newblock \bibinfo{journal}{J. Fluid Mech.} \bibinfo{volume}{912},
  \bibinfo{pages}{A35}.
%Type = Article
\bibitem[{Lohse and Zhang(2020)}]{Lohse2020}
\bibinfo{author}{Lohse, D.}, \bibinfo{author}{Zhang, X.}, \bibinfo{year}{2020}.
\newblock \bibinfo{title}{Physicochemical hydrodynamics of droplets out of
  equilibrium}.
\newblock \bibinfo{journal}{Nat. Rev. Phys.} \bibinfo{volume}{2},
  \bibinfo{pages}{426--443}.
%Type = Article
\bibitem[{Maxworthy(1974)}]{Maxworthy1974}
\bibinfo{author}{Maxworthy, T.}, \bibinfo{year}{1974}.
\newblock \bibinfo{title}{Turbulent vortex rings}.
\newblock \bibinfo{journal}{J. Fluid Mech.} \bibinfo{volume}{64},
  \bibinfo{pages}{227--240}.
%Type = Article
\bibitem[{Mazzino and Rosti(2021)}]{Mazzino2021}
\bibinfo{author}{Mazzino, A.}, \bibinfo{author}{Rosti, M.E.},
  \bibinfo{year}{2021}.
\newblock \bibinfo{title}{Unraveling the secrets of turbulence in a fluid
  puff}.
\newblock \bibinfo{journal}{Phys. Rev. Lett.} \bibinfo{volume}{127},
  \bibinfo{pages}{094501}.
%Type = Article
\bibitem[{McKim et~al.(2020)McKim, Jeevanjee and Lecoanet}]{McKim2020}
\bibinfo{author}{McKim, B.}, \bibinfo{author}{Jeevanjee, N.},
  \bibinfo{author}{Lecoanet, D.}, \bibinfo{year}{2020}.
\newblock \bibinfo{title}{Buoyancy-driven entrainment in dry thermals}.
\newblock \bibinfo{journal}{Q J R Meteorol Soc.} \bibinfo{volume}{146},
  \bibinfo{pages}{415--425}.
%Type = Article
\bibitem[{Middleton(1975)}]{Middleton1975}
\bibinfo{author}{Middleton, J.H.}, \bibinfo{year}{1975}.
\newblock \bibinfo{title}{The asymptotic behaviour of a starting plume}.
\newblock \bibinfo{journal}{J. Fluid Mech.} \bibinfo{volume}{72},
  \bibinfo{pages}{753--771}.
%Type = Article
\bibitem[{de~Rivas and Villermaux(2016)}]{Rivas2016}
\bibinfo{author}{de~Rivas, A.}, \bibinfo{author}{Villermaux, E.},
  \bibinfo{year}{2016}.
\newblock \bibinfo{title}{Dense spray evaporation as a mixing process}.
\newblock \bibinfo{journal}{Phys. Rev. Fluids} \bibinfo{volume}{1},
  \bibinfo{pages}{014201}.
%Type = Article
\bibitem[{Rocco and Woods(2015)}]{Rocco2015}
\bibinfo{author}{Rocco, S.}, \bibinfo{author}{Woods, A.}, \bibinfo{year}{2015}.
\newblock \bibinfo{title}{Dispersion in two-dimensional turbulent buoyant
  plumes}.
\newblock \bibinfo{journal}{J. Fluid Mech.} \bibinfo{volume}{774},
  \bibinfo{pages}{R1}.
%Type = Article
\bibitem[{Shariff and Leonard(1992)}]{Shariff1992}
\bibinfo{author}{Shariff, K.}, \bibinfo{author}{Leonard, A.},
  \bibinfo{year}{1992}.
\newblock \bibinfo{title}{Vortex rings}.
\newblock \bibinfo{journal}{Annu. Rev. Fluid Mech.} \bibinfo{volume}{24},
  \bibinfo{pages}{235--279}.
%Type = Article
\bibitem[{Skvortsov et~al.(2021)Skvortsov, DuBois, Jamriska and
  Kocan}]{Skvortsov2021}
\bibinfo{author}{Skvortsov, A.}, \bibinfo{author}{DuBois, T.C.},
  \bibinfo{author}{Jamriska, M.}, \bibinfo{author}{Kocan, M.},
  \bibinfo{year}{2021}.
\newblock \bibinfo{title}{Scaling laws for extremely strong thermals}.
\newblock \bibinfo{journal}{Phys. Rev. Fluids} \bibinfo{volume}{6},
  \bibinfo{pages}{053501}.
%Type = Article
\bibitem[{Tan et~al.(2019)Tan, Diddens, Mohammed, Li, Versluis, Zhang and
  Lohse}]{Tan2019}
\bibinfo{author}{Tan, H.}, \bibinfo{author}{Diddens, C.},
  \bibinfo{author}{Mohammed, A.A.}, \bibinfo{author}{Li, J.},
  \bibinfo{author}{Versluis, M.}, \bibinfo{author}{Zhang, X.},
  \bibinfo{author}{Lohse, D.}, \bibinfo{year}{2019}.
\newblock \bibinfo{title}{Microdroplet nucleation by dissolution of a
  multicomponent drop in a host liquid}.
\newblock \bibinfo{journal}{J. Fluid Mech.} \bibinfo{volume}{870},
  \bibinfo{pages}{217--246}.
%Type = Article
\bibitem[{Turner(1962)}]{Turner1962}
\bibinfo{author}{Turner, J.}, \bibinfo{year}{1962}.
\newblock \bibinfo{title}{The starting plume in neutral surroundings}.
\newblock \bibinfo{journal}{J. Fluid Mech.} \bibinfo{volume}{13},
  \bibinfo{pages}{356--368}.
%Type = Article
\bibitem[{Uijttewaal(2014)}]{Uijttewaal2014}
\bibinfo{author}{Uijttewaal, W.}, \bibinfo{year}{2014}.
\newblock \bibinfo{title}{Hydrodynamics of shallow flows: application to
  rivers}.
\newblock \bibinfo{journal}{J. Hydraul. Res.} \bibinfo{volume}{52},
  \bibinfo{pages}{157--172}.
%Type = Article
\bibitem[{Villermaux et~al.(2017)Villermaux, Moutte, Amielh and
  Meunier}]{Villermaux2017}
\bibinfo{author}{Villermaux, E.}, \bibinfo{author}{Moutte, A.},
  \bibinfo{author}{Amielh, M.}, \bibinfo{author}{Meunier, P.},
  \bibinfo{year}{2017}.
\newblock \bibinfo{title}{Fine structure of the vapor field in evaporating
  dense sprays}.
\newblock \bibinfo{journal}{Phys. Rev. Fluids} \bibinfo{volume}{2},
  \bibinfo{pages}{074501}.
%Type = Article
\bibitem[{Vitale and Katz(2003)}]{Vitale2003}
\bibinfo{author}{Vitale, S.A.}, \bibinfo{author}{Katz, J.L.},
  \bibinfo{year}{2003}.
\newblock \bibinfo{title}{Liquid droplet dispersions formed by homogeneous
  liquid-liquid nucleation: "the ouzo effect"}.
\newblock \bibinfo{journal}{Langmuir} \bibinfo{volume}{19},
  \bibinfo{pages}{4105--4110}.
%Type = Article
\bibitem[{Vybhav and Ravichandran(2022)}]{Vybhav2022}
\bibinfo{author}{Vybhav, G.R.}, \bibinfo{author}{Ravichandran, S.},
  \bibinfo{year}{2022}.
\newblock \bibinfo{title}{Entrainment in dry and moist thermals}.
\newblock \bibinfo{journal}{Phys. Rev. Fluids} \bibinfo{volume}{7},
  \bibinfo{pages}{050501}.
%Type = Article
\bibitem[{Zhang et~al.(2015)Zhang, Lu, Tan, Bao, He, Sun and Lohse}]{Zhang2015}
\bibinfo{author}{Zhang, X.}, \bibinfo{author}{Lu, Z.}, \bibinfo{author}{Tan,
  H.}, \bibinfo{author}{Bao, L.}, \bibinfo{author}{He, Y.},
  \bibinfo{author}{Sun, C.}, \bibinfo{author}{Lohse, D.}, \bibinfo{year}{2015}.
\newblock \bibinfo{title}{Formation of surface nanodroplets under controlled
  flow conditions}.
\newblock \bibinfo{journal}{PNAS} \bibinfo{volume}{112},
  \bibinfo{pages}{9253--9257}.

\end{thebibliography}





% \end{thebibliography}
\end{document}